\documentclass[twoside]{article}
\usepackage[accepted]{aistats2015}

\usepackage{times}
\usepackage{graphicx} 
\usepackage{subfigure}
\usepackage{natbib}
\usepackage{algorithm}
\usepackage{algorithmic}
\usepackage{hyperref}

\usepackage{color}
\usepackage{multirow}
\usepackage{enumerate}
\usepackage{xspace}
\usepackage{enumitem}
\usepackage{Definitions}

\newcommand{\hide}[1]{}
\newcommand{\xhdr}[1]{\vspace{1.7mm}\noindent{{\bf #1.}}}
\newcommand{\casc}{{\mathbf{t}}}
\newcommand{\alphs}{{\mathbf{A}}}

\newcommand{\netrate}{{\textsc{Net\-Rate}}\xspace}
\newcommand{\netsleuth}{{\textsc{Net\-Sleuth}}\xspace}

\begin{document}

\runningauthor{M. Farajtabar, M. Gomez-Rodriguez, N. Du, M. Zamani, H. Zha, L. Song}

\twocolumn[

\aistatstitle{Back to the Past: Source Identification in Diffusion Networks from Partially Observed Cascades}

\aistatsauthor{Mehrdad Farajtabar$^{*}$ \quad Manuel Gomez-Rodriguez$^{\dagger}$ \quad Nan Du$^{*}$ }
\aistatsauthor{Mohammad Zamani$^{\diamond}$ \quad Hongyuan Zha$^{*}$  \quad Le Song$^{*}$}

\aistatsaddress{$^{*}$Georgia Institute of Technology \quad $^{\dagger}$MPI for Software Systems \quad $^{\diamond}$Stony Brook University} ]


\begin{abstract}

When a piece of malicious information becomes rampant in an information diffusion network, can we identify the \emph{source} node that originally introduced the piece 
into the network and infer the time when it initiated this? 
Being able to do so is critical for cur\-tai\-ling the spread of malicious information, 
and reducing the potential losses incurred. 
This is a very challenging problem since typically only incomplete traces are observed and we need to unroll the incomplete traces into the past in order to pinpoint 
the source. 
In this paper, we tackle this problem by deve\-lo\-ping a two-stage framework, which first learns a continuous-time diffusion network model based on historical diffusion traces 
and then identifies the source of an incomplete diffusion trace by maximizing the likelihood of the trace under the learned model. 
Experiments on both large synthetic and real-world data show that our framework can effectively ``go back to the past'', and pinpoint the source node and its initiation time 
significantly more accurately than previous state-of-the-arts. 
\end{abstract}

\section{INTRODUCTION}
\label{sec:introduction}

On September 2014, a collection of hundreds of private pictures from various celebrities, mostly consisting of women and often containing nudity, were posted online, and later 
disseminated by users on websites and social networks such as Imgur\footnote{http://imgur.com/}, Reddit\footnote{http://www.reddit.com/} and Tumblr\footnote{https://www.tumblr.com/}~\citep{time}. 
After quite some efforts in ma\-nual tracing of the diffusion paths, investigators found that the imageboard 4chan\footnote{http://www.4chan.org/} was the culprit site where the photos were originally posted 
on August 31, even though the photos had been taken down from the site soon after their post.
This leakage of private pictures has touched off a larger world-wide discussion and debate on the state of privacy and civil liberties on the Internet~\citep{nyt}.

Can we automatically pinpoint the identity of such malicious information sources, as well as the time when they first posted the malicious information, given historical incomplete diffusion traces? 
Solving this source identification problem is of outstanding interest in many scenarios~\citep{lappas2010finding}. For example, finding people that originate rumors may reduce disinformation, 
identifying patient zeros in disease spreads may help to understand and control epidemics, or inferring where a trojan or computer worm is initially released may increase reliability of computer 
networks.

{\bf Related Work.} The problem of finding the source of a diffusion trace, also called \emph{cascade}, has not been studied until very recently~\citep{lappas2010finding, 
shah2010detecting, prakash2012, pinto2012locating}.
However, most previous work assumes that a complete steady-state snapshot of the cascade is observed, in other words, we know \emph{which} nodes got \emph{infected} but  
\emph{not when} they did so.
Moreover, previous work uses discrete-time sequential propagation models such as the independent cascade model~\citep{kempe03maximizing} or the discrete version of the SIR 
model~\citep{bailey75mathematical}, which are difficult to estimate accurately from real world data~\citep{manuel11icml, nandu12nips, DuSonWooZha13, ZhoZhaSon13, ZhoZhaSon13b}.

Only very recently,~\cite{pinto2012locating} consider a fairly general continuous-time model and assume that only a small fraction of sparsely-placed nodes are observed 
and, if infected, their infection time is observed. Unfortunately, their approach requires the distance between observed nodes to be large because they approximate 
the infection times by Gaussian distributions using the central limit theorem. Since this is easily violated in real social and information networks~\citep{backstrom2012four},
we find its performance on this type of networks underwhelming, as shown in Section~\ref{sec:experiments}.
\begin{figure}[t]
  \centering
  \includegraphics[width=0.50\textwidth]{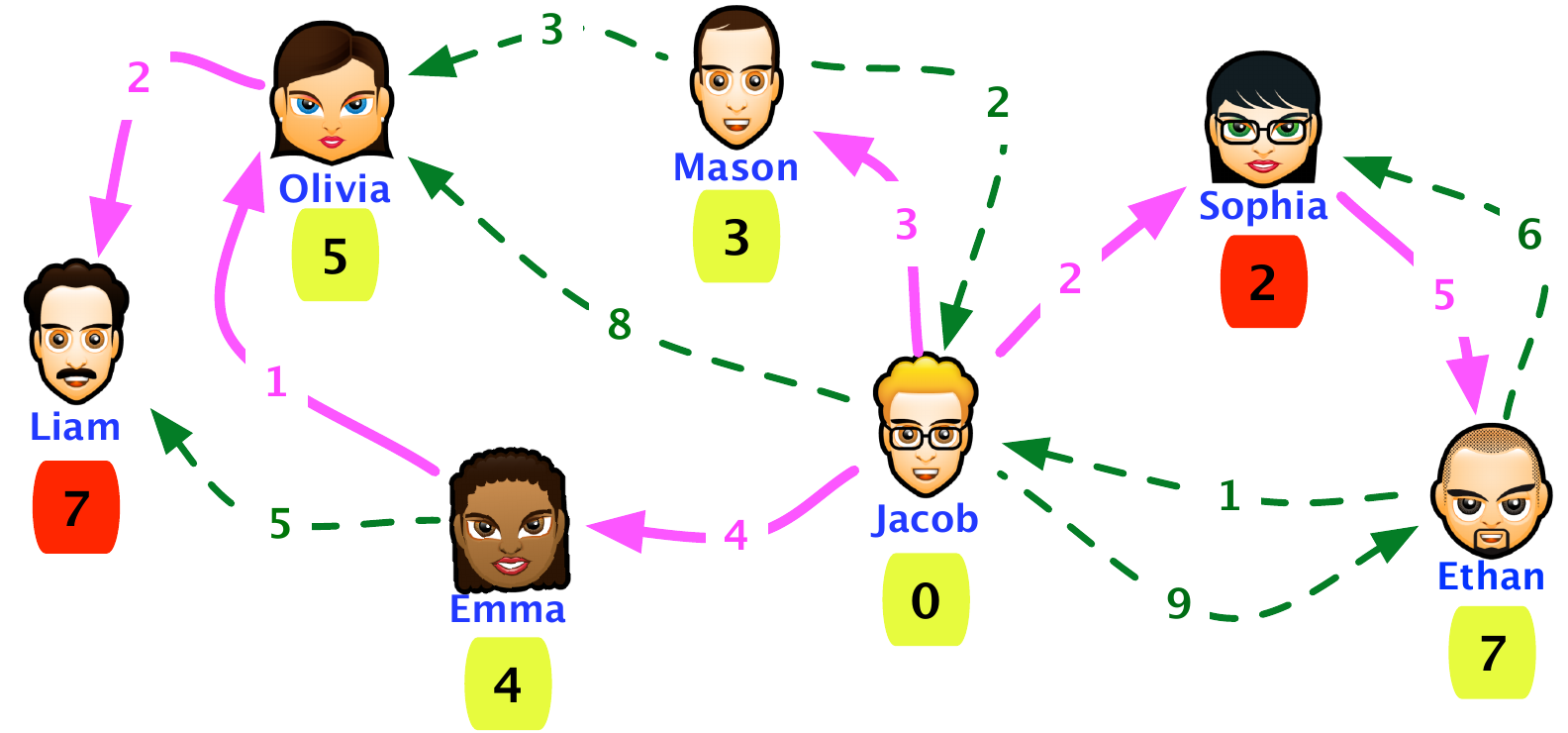} 
  \caption{Spread of a rumor in a social network. Each edge weight is the time it took for a rumor to pass along the edge. Solid magnet edges indicate the actual path through which 
  the rumor spreads. Green dashed edges are alternative ways in which the rumor could have spread. The infection times of Sophia and Liam are observed (red squares); the
  infection times of the remaining nodes are hidden (yellow squares). How can an algorithm find that Jacob was the person who initiated the rumor? 
  \label{fig:social-network}} 
\end{figure}

%
%

{\bf Challenges.} Previous approaches failed suc\-cess\-fully address\- several challenges of the source identification problem, which we illustrate next using 
a toy example, shown in Figure~\ref{fig:social-network}. 

\emph{--- Partially observed infections.} It has become di\-ffi\-cult, if not impossible, to collect complete diffusion traces, and track each individual infection in online social and information networks.
This problem is exa\-cer\-bated by the need to develop methods that can provide outputs in (almost) real-time.
For example, Spinn3r\footnote{http://spinn3r.com/} crawls only a subset of the blogs periodically; Twitter's streaming API provides a small percentage (1\%) of the full stream of tweets~\citep{morstatter2013sample}; Facebook 
users typically keep their activity and profiles private~\citep{sadikov11cascades}. It is thus ne\-ce\-ssary to develop methods that are robust to missing data~\citep{chierichetti11,kim2011network,sadikov11cascades}.
Our toy example illustrates this challenge by considering the infection times of Liam and Sophia to be observed and all other infection times to be missing (hidden or unobserved). 

\emph{--- Unknown infection start time.} In most real-world scenarios, the exact time when a piece of malicious information starts spreading is unknown, and thus the observed infection times have only 
relative meaning.
In our toy example, we know Liam got infected $5$ time units later than Sophia but we do not actually observe how much time has passed between Jacob'{}s infection, which triggered the spread, and Sophia'{}s infection.

\emph{--- Uncertain transmission delay.} The spread of information over social and information networks is a stochastic process. Therefore, we need to consider probabilistic transmission models to 
capture the uncertainty. 
For example, our toy example illustrates the spread of a particular rumor and therefore considers a set of fixed edge delays (\eg, the rumor took $5$ time units to spread from Sophia to Ethan). 
However, the edge delays are stochastic and possibly different for eve\-ry particular rumor (\eg, a different rumor can take more or less than $5$ time units to spread from Sophia to Ethan). 
The edge delay densities, or transmission densities, may depend on parameters like the content of the rumor or the users'{} influence.

\emph{--- Unknown infection path.} In large real world networks, we will often encounter a large number of potential paths that may explain the spread of a rumor from a source node and any other 
node in the network.
In fact, the set of potential paths increases exponentially with network size and network density and even simply counting the number of paths requires non-trivial methods~\citep{influmax12icml,multitree12icml,du2013scalable}.
For example, in our toy example, Liam can become aware of the rumor through either Olivia or Emma.

{\bf Our Approach.} To tackle these challenges, we propose a two-stage scalable framework: we first learn a continuous-time diffusion network model based on his\-to\-ri\-cal diffusion traces and then identify the source of an incomplete diffusion trace by maximizing its likelihood under the learned model. 
The key idea of our framework is to view the problem from the perspective of graphical models, and cast the problem as a maxi\-mum likelihood estimation problem, for which we find optimal solutions 
very efficiently using an importance sampling approximation to the objective and an optimization procedure that exploits the structure of the problem. 
Additionally, for networks with exponentially distributed edge transmission densities, used previously for modeling information propagation~\citep{manuel11icml}, we show that the objective is a 
piece-wise unimodal function with respect to the source'{}s infection time and develop a more efficient search procedure.



For both synthetic and real-world data, we show that the framework can effectively ``travel back to the past", and pinpoint the source node and its infection time significantly more accurately 
than other  methods. 

\section{OUR FRAMEWORK}
\label{sec:model}
\setlength{\abovedisplayskip}{3pt}
\setlength{\abovedisplayshortskip}{1pt}
\setlength{\belowdisplayskip}{3pt}
\setlength{\belowdisplayshortskip}{1pt}
\setlength{\jot}{2pt}
\setlength{\floatsep}{2ex}
\setlength{\textfloatsep}{2ex}
Our framework for solving the source identification problem consists of two main stages: it first learns a continuous-time diffusion network model based on historical diffusion traces, 
and then identifies the source of an incomplete diffusion trace and its initiation time by maximizing the its likelihood under the learned model. We start our exposition by 
revisiting the continuous-time generative model for cascade data in social networks introduced in~\citet{manuel11icml,du2013scalable}.

\subsection{Continuous-Time Model for Cascades}

Given a \emph{directed} contact network, $\Gcal = (\Vcal,\Ecal)$ with $N$ nodes, a diffusion process begins with an infected source node $s$ initially adopting certain \emph{contagion} 
(idea, rumor or malicious piece of information) at time $t_s$. The contagion is transmitted from the source along her out-going edges to 
their direct neighbors. Each transmission through an edge entails a \emph{random} spreading time, $\tau$, drawn from a density over time, $f_{ji}(\tau ; \alpha_{ji})$, parametrized 
by a transmission rate $\alpha_{ji}$.
Then, the infected neighbors transmit the contagion to their respective neighbors, and the process continues.
We assume\- transmission times are independent and nonnegative, in other words, a node cannot be infected by a node infected later in time; $f_{ji}(\tau ; \alpha_{ji}) = 0$ if $\tau < 0$.
Moreover, an infected nodes remain infected for the entire diffusion process. Thus, if a node $i$ is infected by multiple neighbors, only the neighbor that first infects node $i$ will be the 
\emph{true parent}.
%
%

The temporal traces left by diffusion processes are often called \emph{cascades}.
A cascade $\casc$ is an $N$-dimensional vector $\casc := (t_1,\ldots,t_N)$ recording the times when nodes are infected, if so, \ie, $t_i \in [0, \infty]$, where $T$ is the observation
window cut-off and $\infty$ denotes nodes that did not get infected during the observation window.
However, as noted above, in many scenarios, we only observe a subset of the infected nodes, $\Ocal$, while the state of all other nodes, $\Hcal$, is hidden (we assume the source node $s \in \Hcal$). Our aim is then to find the source of a cascade $s$ from the infection times $\{ t_j \}_{j \in \Ocal}$ of the subset of infected nodes $\Ocal$.
Figure~\ref{fig:social-network} illustrates the observed data. 

\subsection{Cascade Likelihood}

According to the conditional independence relation proposed in the continuous-time model for cascades, the complete likelihood of a cascade $\casc$ (for both observed and hidden 
nodes) factorizes as
%
\begin{align}  \label{eq:full-cascade-log-likelihood}
  p(\casc | t_s) = \prod_{i \in \Ocal \cup \Hcal} p\rbr{t_i |\{t_j\}_{j \in \pi_i}}
\end{align}
where $\pi_i$ is the set of parents of $i$ defined by the directed graph $\Gcal$. For the infected nodes, \citet{manuel11icml} showed that the likelihood can be further 
written as
\begin{align*}
  p\rbr{t_i |\{t_j\}_{j \in \pi_i}} = \prod_{j \in \pi_i} S(t_i-t_j; \alpha_{ji}) \sum_{l \in \pi_i} H(t_i-t_l; \alpha_{li}),
\end{align*}
where $S_{ji}(\tau;\alpha_{ji}) = 1-F_{ji}(\tau; \alpha_{ji})$ is the survival function, $F_{ji}(\tau;\alpha_{ji}) = \int_0^{\tau} f_{ji}(t;\alpha_{ji}) dt$ is the cumulative distribution function, and $H_{ji}(\tau;\alpha_{ji}) = \frac{f_{ji}(\tau; \alpha_{ji})}{S_{ji}(\tau;  \alpha_{ji})}$ is the hazard function, or instantaneous infection rate. 
%
%
We will focus on the Weibull family of distributions $f_{ji}(\tau;\alpha_{ji})$ since they have been shown to fit well real world diffusion data~\citep{DuSonWooZha13}. In this case, 
\begin{align*}
  f_{ji}(\tau; \alpha_{ji}) = \frac{k\tau^{k-1}}{\alpha_{ji}^k}e^{-\rbr{\frac{\tau}{\alpha_{ji}}}^k},~S_{ji}(\tau; \alpha_{ji}) = e^{-\rbr{\frac{\tau}{\alpha_{ji}}}^k},
\end{align*}
where $k$ is a hyperparameter controlling the shape of the density. This family includes many well-known special cases, such as the exponential or Rayleigh distributions, which have also been used to model information pro\-pa\-ga\-tion over information networks~\citep{manuel10netinf,DuSonWooZha13}.

Unfortunately, to use Eq.~\ref{eq:full-cascade-log-likelihood}, all infected nodes in a cascade need to be fully observed. 
If we only observe a subset $\Ocal$ of the infected nodes, the likelihood of the incomplete cascade is computed as follows,
\begin{align} \label{eq:partial-cascade-log-likelihood}
  p(\cbr{t_i}_{i \in \Ocal} | t_s)
  &=\int_{\Omega} p\rbr{\casc | t_s} \prod_{j \in \Hcal}\, dt_j \nonumber \\
  &= \int_{\Omega} \prod_{i \in \Ocal \cup \Hcal} p\rbr{t_i | \{t_j\}_{j \in \pi_i}} \prod_{j \in \Hcal}\, dt_j,
\end{align}
which essentially marginalize out the time for all hidden nodes $\Hcal$ over a product space
$\Omega := [t_s,\infty)^{|\Hcal|}$. For simplicity of notation, we will omit the domain of the integration in the remainder of the paper. 

The computation of the incomplete likelihood is a difficult high dimensional integration problem for continuous variables. We will address this technical challenge using\- importance sampling in Section~\ref{sec:sampling-infection-times}.

\subsection{Learning Diffusion Networks}

%
Our framework relies on the assumption that it is possi\-ble to record a sufficiently large number of historical cascades, $\Ccal$, in order to discover the existence of all nodes in
the network, to infer the network structure as well as the model parameters, $\cbr{\alpha_{ji}}$. We note that it is not necessary to record cascades that cover all nodes and edges, but each cascade has to be \emph{fully} observed to a suffi\-cient\-ly large time period. Furthermore, all cascades collectively need to cover the entire diffusion network. Under the precise conditions stated in~\citep{daneshmand14netrate}, one can infer the parameters of the continuous time model using an $\ell_1$-regularized maximum likelihood estimation procedure.


\subsection{Cascade Source Identification Problem} 

Given a learned diffusion model, our aim is to find the source node $s$ of an incomplete cascade, such that the log-likelihood of the incomplete cascade is maximized. Thus, we 
aim to solve
\begin{equation} \label{eq:full-opt-problem}
 s^* = \argmax_{s \in \Hcal}\,  \max_{t_s \in (-\infty, \min_{i \in \Ocal} t_i)}  p(\cbr{t_i}_{i \in \Ocal}|t_s),
 \end{equation}
where $p(\cbr{t_i}_{i \in \Ocal}| t_s)$ is defined in Eq.~\ref{eq:partial-cascade-log-likelihood}, and we assume that 
$t_s < \min_{i \in \Ocal} t_i$. 
If we observe several independent incomplete cascades $\Dcal$, all triggered by the same source node, we will maximize their joint likelihood
\begin{equation} \label{eq:full-opt-problem-several-cascades}
 s^* = \argmax_{s \in \Hcal}\, \prod_{c \in \Dcal} \left( \max_{t^c_s \in (-\infty, \min_{j \in \Ocal} t^c_j)} \mathcal{L}_c \right),
 \end{equation}
 where $\mathcal{L}_c :=  p(\cbr{t^c_j}_{j \in \Ocal}| t_s)$.
 %
In the following sections, we will design algorithms to efficiently optimize the above objective and present experimental evaluations.  

\vspace{-3mm}
\section{APPROXIMATE OBJECTIVE FUNCTION}
\label{sec:method}
\vspace{-3mm}
There remain two technical challenges to solve to make our framework useful in practice.
First, the likelihood of incomplete cascades, given by Eq.~\ref{eq:partial-cascade-log-likelihood}, is a difficult high dimensional integration problem 
over a continuous domain. 
We overcome this difficulty by an approximation algorithm, based on importance sampling, which will greatly simplify the integration. 
Second, the inner-loop maximization over the source timing in Eq.~\ref{eq:full-opt-problem-several-cascades} is non-convex. 
We solve this by designing an effi\-cient algorithm, which finds the global maximum by exploiting the piece-wise structure of the problem.


\vspace{-3mm}
\subsection{Importance Sampling}
\vspace{-3mm}

Since an analytical evaluation of the integral in Eq.~\ref{eq:partial-cascade-log-likelihood} is intractable, we turn to a Monte Carlo approximation. To do so, in principle, 
we need to draw samples from the posterior distribution of latent variables, $p(\{t_i\}_{i \in \Hcal} |t_s, \{t_i\}_{i \in \Ocal})$, given the source time $t_s$ and the times of 
the observed nodes, $\{t_i\}_{i\in \Ocal}$. 
However, it is very challenging to sample from this posterior distribution, and we will instead address the problem by designing an efficient importance sampling approach.  

More specifically, we first introduce a set of \emph{auxiliary random variables} $\cbr{ \eta_i }_{i \in \Ocal}$, where each variable corresponds to one observed infected node, 
with an arbitrary joint probability distribution $\tilde{q}(\cbr{ \eta_i }_{i \in \Ocal})$. 
In the next section, we will briefly discuss how $\tilde{q}$ is chosen.
Then, given the auxiliary distribution we have
%
\begin{align}
  & p(\cbr{t_i}_{i \in \Ocal} | t_s) =  
  \int  p(\cbr{t_i}_{i \in \Ocal \cup \Hcal } | t_s) \prod_{i \in \Hcal} d t_i \nonumber \\
  = &  \int  p(\cbr{t_i}_{i \in \Ocal \cup \Hcal } | t_s)  \tilde{q}(\cbr{ \eta_i }_{i \in \Ocal}) \prod_{i \in \Hcal} d t_i   \prod_{i \in \Ocal} d \eta_i. \label{eq:auxiliary}
\end{align}

Second, we introduce the \emph{proposal distribution} for importance sampling on the auxiliary and hidden variables, 
$ q(\cbr{ \eta_i }_{i \in \Ocal}, \cbr{t_i}_{i \in \Hcal} )$. Then, the integral becomes
\begin{align}
  p(\cbr{t_i}_{i \in \Ocal} | t_s) 
  =  & \int \frac{p(\cbr{t_i}_{i \in \Ocal \cup \Hcal } | t_s)  \tilde{q}(\cbr{ \eta_i }_{i \in \Ocal})} {q(\cbr{ \eta_i }_{i \in \Ocal}, \cbr{t_i}_{i \in \Hcal}) }  \nonumber \\
  & q(\cbr{ \eta_i }_{i \in \Ocal}, \cbr{t_i}_{i \in \Hcal} ) \prod_{i \in \Hcal} d t_i   \prod_{i \in \Ocal} d \eta_i \nonumber\\
  \approx & \frac{1}{L} \sum_{l=1}^{L} \frac{p(\cbr{t_i}_{i \in \Ocal}, \cbr{t_i^l}_{i \in \Hcal } | t_s)  \tilde{q}(\cbr{ \eta_i^l }_{i \in \Ocal})} {q(\cbr{ \eta_i^l }_{i \in \Ocal}, \cbr{t_i}_{i \in \Hcal}) } \nonumber\\
  \triangleq & \phi_L(t_s), \label{eq:sampling}
\end{align}
where we draw $L$ samples from $q(\cbr{ \eta_i }_{i \in \Ocal}, \cbr{t_i}_{i \in \Hcal}) $ to approximate the integral.
Now, we have an approximation to $\Lcal_c$. Next, we explain how to choose the \emph{proposal} and the \emph{auxiliary} distributions.

\subsection{Choice of Proposal Distributions}
\label{sec:sampling-infection-times}
%
We define our proposal distribution using the forward\- generative process of the cascades. Our propo\-sal dis\-tri\-bution $q(\cbr{ \eta_i }_{i \in \Ocal}, \cbr{t_i}_{i \in \Hcal})$ 
will sample cascades from the learned continuous diffusion network model with $s$ as the source set.
One of the \-interesting properties of this proposal distribution is that many terms involving the latent variables in Eq.~\ref{eq:sampling} will be canceled out and hence the formula will 
become simpler.  

We remind the reader that  the independent cascade model has a useful shortest-path property~\citep{du2013scalable}, which allows us to sample the parents'{} infection times, 
$\{t_i \}_{i \in \pi_j}$, for each node $j$ efficiently for different source infection times $t_s$. 
More specifically, we first sample a set of transmission times $\{ \tau_{uv} \}_{(u,v) \in \Ecal}$, one per edge, independent of each other. Then, the time $t_i$ taken to infect a node 
$i$ is simply the length of the shortest path in $\Gcal$ from the source $s$ to node $i$, where the edge weights correspond to the associated transmission times.
Let $\Qcal_i(s)$ be the collection of directed paths in $\Gcal$ from the source $s$ to node $i$, where each path $q\in \Qcal_i(s)$ contains 
a sequence of directed edges $(j,m)$, and assume the source node is infected at time $t_s$, then we obtain variable $t_i$ via
\begin{equation} \label{eq:shortest-path}
 t_i = g_i\rbr{\{\tau_{jm}\}_{(j,m)\in \Ecal} | s}
 := \min_{q \in \Qcal_i(s)} \sum\nolimits_{(j,m)\in q} \tau_{jl},
\end{equation}
where $g_i(\cdot)$ is the value of the shortest-path. 

This above relation is key to speed up the evaluation of the sampled likelihood in Eq.~\ref{eq:sampling} for different $t_s$ values. 
First, the sampled transmission times $\tau_{uv}$ are independent and thus can be sampled in parallel.
Second, we can reuse the sampled transmission times $\tau_{uv}$ for different $t_s$ values and sources $s$, since the transmission times are independent of $t_s$ and $s$. 
We only need to compute the infection time $t_i$ for each node using $t_s=0$, and then for a different value of $t_s$, the infection time is just an offset by $t_s$. 
%
%
Third, the likelihood of a sampled cascade  $(\cbr{ \eta_i^l }_{i \in \Ocal}, \cbr{t_i^l}_{i \in \Hcal})$ for $l=1, \ldots, L$ can be simply computed using Eq.~\ref{eq:full-cascade-log-likelihood} 
as $p(\cbr{\eta_i^l}_{i \in \Ocal}, \cbr{t_i^l}_{i \in \Hcal } | t_s)$, which is independent of the actual value of $t_s$ and depends only on the identity of the source node $s$.

\begin{algorithm*}[t]
\caption{Our source detection algorithm} \label{alg:overall-algorithm}
\begin{algorithmic}
\REQUIRE $\Ccal, \Dcal, L$
\STATE Infer transmission rates $\alphs$ from $\Ccal$ using~\cite[Algorithm 1]{daneshmand14netrate}.
\STATE Sample $L$ sets of transmission times $\{\tau_{ij}\}_{(i,j)\in\Ecal}$.
\STATE Compute infection times $\hat{t}^{l}_{i \in \Vcal},\, l=1,\ldots,L$ assuming $t_s = 0$ using Eq.~\ref{eq:shortest-path}.
\STATE Compute change points: $t_i - \hat{t}^{l}_{j}, i \in \Dcal, j \in \pi_i, l=1, \ldots, L$ and $t_i - \hat{t}^{l}_j, j \in \Mcal, i \in \pi_j \cap \Ocal, l=1, \ldots, L$.
\FOR{$i \in \Mcal$}
  \STATE $t^{*}_{i} = \argmax_{t_s} \phi_L(t_s)$ (using line search method or Lemma~\ref{lemma2} in each piece)
\ENDFOR
\STATE $s^{*} = \argmax_{i \in \Mcal} \phi_L(t^{*}_{i})$
\STATE $t_{s^{*}} = \max_{i \in \Mcal} \phi_L(t^{*}_{i})$
\end{algorithmic}
\vspace{-1mm}
\end{algorithm*}

\subsection{Choice of Auxiliary Distribution}

The auxiliary distribution $\tilde{q}(\cbr{ \eta_i}_{i \in \Ocal})$ is chosen to be equal to $p(\cbr{ \eta_i}_{i \in \Ocal} | \cbr{t_i}_{i \in \Hcal})$. In other words, our auxi\-lia\-ry distribution will 
simply sample cascades from the learned con\-ti\-nuous diffusion network model with $\Hcal$ as the source set. 
Here, it is is easy to see that
$\int  p(\cbr{ \eta_i }_{i \in \Ocal}|\cbr{t_i }_{i \in \Hcal}) \prod_{i \in \Ocal} d \eta_i = 1$.
With the above choices for the proposal and auxiliary distribution, we can greatly simplify the approximate  likelihood in Eq.~\ref{eq:sampling} into 
\begin{align} \label{eq:sampling-simplified}
  \phi_L(t_s) = 
  \frac{1}{L} \sum_{l=1}^{L} &{\prod_{i \in \Ocal} p(t_i | \cbr{t_j^l}_{j \in \pi_i \setminus \Ocal}, \cbr{t_j}_{j \in \pi_i \cap \Ocal})}  \\
  &\prod_{i \in \Mcal} \frac{p(t_i^l | \cbr{t_j^l}_{j \in \pi_i \setminus \Ocal}, \cbr{t_j}_{j \in \pi_i \cap \Ocal})}
  {p(t_i^l | \cbr{t_j^l}_{j \in \pi_i \setminus \Ocal}, \cbr{\eta_j^l}_{j \in \pi_i \cap \Ocal} )}, \nonumber
\end{align}
where $\Mcal$ is the set of hidden nodes with observed variables as parents,
\begin{align}
  \Mcal:=\cbr{i\in \Hcal | \pi_i \cap \Ocal \neq \emptyset},
\end{align}
which is typically much smaller than the overall set of hidden nodes.
It is noteworthy that, under mild regu\-larity conditions, the Monte Carlo approximation of the integral will converge to the true value with sufficient number of samples. 
However, a clever choice of the proposal distribution makes the convergence faster and the computation more efficient.

\section{MAXIMIZE OBJECTIVE FUNCTION}

%
Our objective function, given by Eq.~\ref{eq:full-opt-problem-several-cascades}, consists of an inner and an outer maximization.
In the inner maximization, we leverage the Monte Carlo sample approxi\-ma\-tion and solve
\begin{equation}
\label{eq:inner-max-with-sampling}
 \max_{t_s}~~\phi_L(t_s).
\end{equation}
In the outer maximization, we rank all possible source nodes, $s$, in terms of their best starting time $t_s$, which is the solution to the inner maximization, and then select the top source node in
the ranking as our optimal source, $s^{*}$.

The outer maximization is straightforward, however, the inner maximization, which consists of finding the optimal $t_s$ that maximizes $\phi_L(t_s)$, defined 
in Eq. \ref{eq:sampling-simplified}, may seem difficult at first. 
Although it is a 1-dimensional problem, the objective function is piece-wise continuous and non-convex with respect to $t_s$. This is because by increasing (or decreasing) $t_s$, 
the parent-child relation between nodes may change.
However, there are two key properties of $\phi_L(t_s)$, which allow us to carry out the optimization efficiently. 
First, $\phi_L(t_s)$ is piece-wise continuous and the number of such pieces increases as $O(L\Delta N)$,  \ie, linearly in the number of Monte Carlo samples, the number of observed nodes,
and the maximum in-degree, $\Delta$, of the observed nodes. 
Second, within each piece, the maximum of the function can be found efficiently.

\subsection{Finding Each Continuous Piece}

%
In this section, we aim to efficiently find all the change points $t_{s_i}$ in the approximated likelihood $\phi_L(t_s)$, given by Eq.~\ref{eq:sampling-simplified}. In other words, we will 
efficiently find the left and right end points of each of its continuous pieces.
Here, we assume there is a directed path in $\Gcal$ from the source $s$ to each of the observed infected nodes $\Ocal$, otherwise, it cannot be a source for those 
nodes, trivially.

The key idea to finding all change points is realizing that each piece in Eq.~\ref{eq:sampling-simplified} corresponds to a different \emph{feasible} parents-child configuration. Here, by feasible 
parents, we mean parents that get infected earlier than the child and thus are temporally plausible.
%
%
More specifically, given a source $s$, Eq.~\ref{eq:sampling-simplified} is composed of three types of terms:
$p(t_i | \cbr{t_j^l}_{j \in \pi_i \setminus \Ocal}, \cbr{t_j}_{j \in \pi_i \cap \Ocal})$ and $p(t_i^l | \cbr{t_j^l}_{j \in \pi_i \setminus \Ocal}, \cbr{t_j}_{j \in \pi_i \cap \Ocal})$, which depend on the source time 
value $t_s$, as we will realize shortly, and thus are responsible for the change point values $t_{s_i}$, and 
$p(t_i^l | \cbr{t_j^l}_{j \in \pi_i \setminus \Ocal}, \cbr{\eta_j^l}_{j \in \pi_i \cap \Ocal} )$, which does not depend on $t_s$, because both $\cbr{ \eta_i^{l}}_{i \in \Ocal}$ and 
$\cbr{ t_i^{l}}_{i \in \Mcal}$ are sampled, $t_s$ equally shifts all sampled times and its likelihood is time shift invariant.
%
%
%
%
%
Based on the structure of the first two type of terms, it is easy to show that at each change point $t_{s_i}$, there is a node $j \in \Ocal \cup \Mcal$, observed or hidden, that changes its 
set of feasible temporally plausible parents, \ie, a parent of one observed or hidden\- node becomes (stops being) a feasible parent at time $t_{s_i}$.
Therefore, it is clear that there are $O(L \Delta N)$ change points, where $\Delta$ is the maximum in-degree of nodes. Next, we describe a procedure to find all change points 
efficiently.

\xhdr{Efficient Change Point Enumeration} We start by setting $t_s = 0$ and computing the infection time, denoted as $\hat{t}^{l}_{j}$, for each hidden node $j \in \Mcal$ and rea\-li\-zation 
$l$ using the shortest path property described in Section~\ref{sec:sampling-infection-times}. 
Then, we find the change points in which an observed node $i \in \Ocal$ looses feasible parents by computing the time difference $t_i - \hat{t}^{l}_{j}, j \in \pi_i \backslash \Ocal, l=1, \ldots, L$, and 
the change points in which a hidden node $j \in \Mcal$ earns feasible parents by computing the time differences $t_i - \hat{t}^{l}_j, i \in \pi_j \cap \Ocal, l=1, \ldots, L$. If a time difference
is negative, we skip it, since the associated parent will never (always) be feasible, independently of the $t_s$ value.

Additionally, we can compute $\phi_L(.)$ efficiently for each change point $t_{s_i}$,
since at each change point $t_{s_i}$, we will only need to revaluate the corresponding terms to the node $i \in \Ocal \cup \Mcal$ that changes its set of feasible parents.
In the case of exponential transmission likelihoods, once we have computed the likelihood at each change point $t_{s_i}$, we can re-evaluate it at any time $t \in [t_{s_i}, t_{s_{i+1}})$, by multiplying 
the corresponding terms in the approximated likelihood by $e^{t-t_{s_i}}$.

\vspace{-3mm}
\subsection{Maximizing within Each Piece}

%
%

%
Once we have delimited each piece of the approximate likelihood given by Eq.~\ref{eq:sampling-simplified}, we can find the times $t_s$ that maximize the likelihood in each piece 
efficiently, using well-known line-search procedures for one-dimensional continuous function, such as the forward-backward method, the golden section method or the Fibonacci 
method~\citep{luenberger1973introduction}. 
%
However, in the case of exponential transmission likelihoods, we can perform the maximization step even more efficiently.
\begin{figure}[t]
  \centering
  \includegraphics[width=0.5\textwidth]{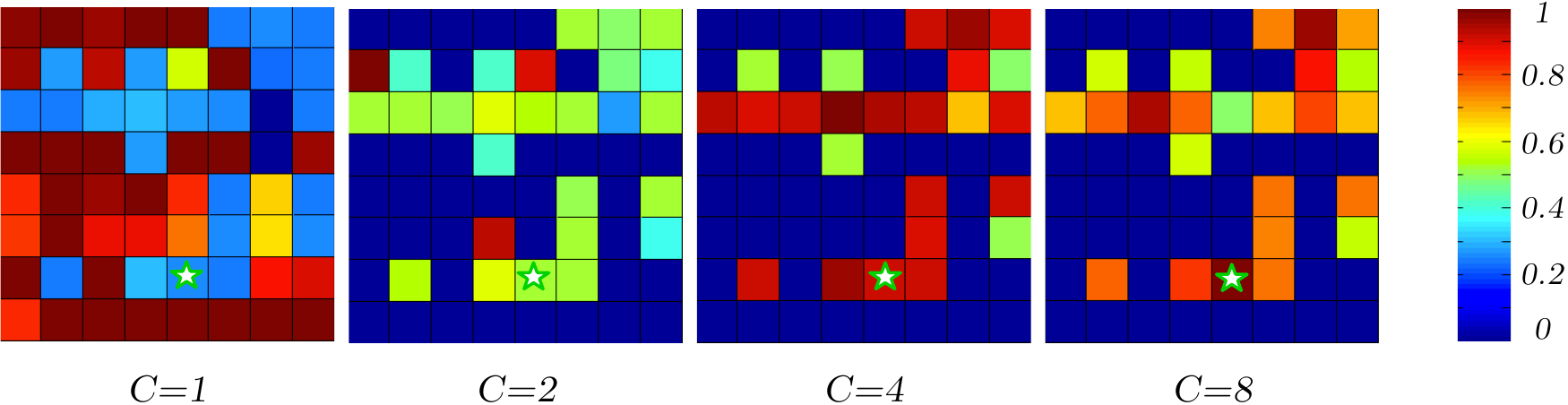} 
  \caption{Evolution of the proposed method with respect to the number of cascades.} \label{fig:grid}
\end{figure}

{\bf Exponential Transmission.} 
%
We start by realizing that, in the case of exponential transmission functions, the approximate likelihood given by Eq.~\ref{eq:sampling-simplified} can be expressed as
\begin{equation}
\begin{split}
\phi_L(t_s) 
=&\sum_{l=1}^{L} \gamma_l e^{\beta_l t_s},
\end{split}
\end{equation}
where $\gamma_l > 0$ and $\beta_l$ are independent of $t_s$. Then, we can prove that each piece of the approximate likelihood is unimodal (proven in Appendix~\ref{app:lemma1-proof}):
%
%
\begin{lemma} \label{lemma1}
$\phi_L(t_s)$ is uni-modal in $ t_{s_i} < t_s < t_{s_{i+1}}$.
\end{lemma}
%
%
Now, we can find the maximum of $\phi_L(.)$ by only evaluation of the function on a sequence of points (proven in Appendix~\ref{app:lemma2-proof}):
\begin{lemma} \label{lemma2}
The maximum point of $\phi_L(.)$ can be found within $\epsilon$-neighborhood of $t_s^*$ with only $2\log(\frac{ t_{s_{i+1}}- t_{s_{i}}}{\epsilon})/ \log(3/2)$ evaluations of $\phi_L(.)$.
%
%
%
%
%
%
\end{lemma}
Furthermore, by utilizing golden section search~\citep{kiefer1953sequential}, one can further reduce the complexity of finding the optimum point to $\log(\frac{ t_{s_{i+1}}- t_{s_{i}}}{\epsilon})/ \log(1.618)$ evaluations.
%

%
We summarize the overall algorithm in Algorithm~\ref{alg:overall-algorithm}.
%

\section{EXPERIMENTS}  
\label{sec:experiments}
 We evaluate the performance of our method on: \emph{(i)} synthetic networks that mimic the structure of social networks and \emph{(ii)} real networks inferred from a large 
cascade dataset, using a well-known state-of-the-art network inference method~\citep{manuel11icml}. 
We show that our approach discovers the true source of a cascade or set of cascades with surprisingly high accuracy in synthetic networks and quite often in real networks, 
given the difficulty of the problem, and significantly outperforms several baselines and two state of the art methods~\citep{prakash2012,pinto2012locating}.
Appendix~\ref{app:experiments} provides additional experimental results.

\subsection{Experiments on Synthetic Data} \label{sec:experiments-synthetic}
%
{\bf Experimental Setup.} We generate three types of Kronecker networks~\citep{leskovec2010kronecker}: 
(\emph{i}) core-periphery networks (parameter matrix: [0.9 0.5; 0.5 0.3]), which mimic the information diffusion traces in real world networks~\citep{manuel10netinf},
(\emph{ii}) random networks ([0.5 0.5; 0.5 0.5]), typically used in physics and graph theory~\citep{easley2010},
and (\emph{iii}) hierarchical networks ([0.9 0.1; 0.1 0.9])~\citep{clauset08hierarchical}. 
We then set the pairwise transmission rates of the edges of the networks by drawing samples from $\alpha \sim U(10, 5)$.  
For each type of Kronecker network, we generate $10$ networks with $256$ nodes and $512$ edges. 
Finally, for each network, we generate a set of cascades from ten different random sources $s^{*}$.
Since we are interested in detecting source nodes of \emph{large} cascades, we only consider source nodes that triggered at least ten 
\emph{large} cascades out of $100$ simulated cascades.
Given the size of the networks we experiment with, we consider a cascade to be \emph{large} if it contains more than $40$ nodes.
Our aim is then to find the source of a \emph{large} cascade or small set of \emph{large} cascades from the infection times of a small (unknown) 
fraction of all infected nodes. In all the following experiments the sample size is 400 and 10\% of the infected nodes are observed except 
when it is explicitly mentioned.
\begin{figure*}[t]
  \centering
  \small
  \setlength{\tabcolsep}{1pt}
  \begin{tabular}{cccc}
    \includegraphics[width=0.245\textwidth]{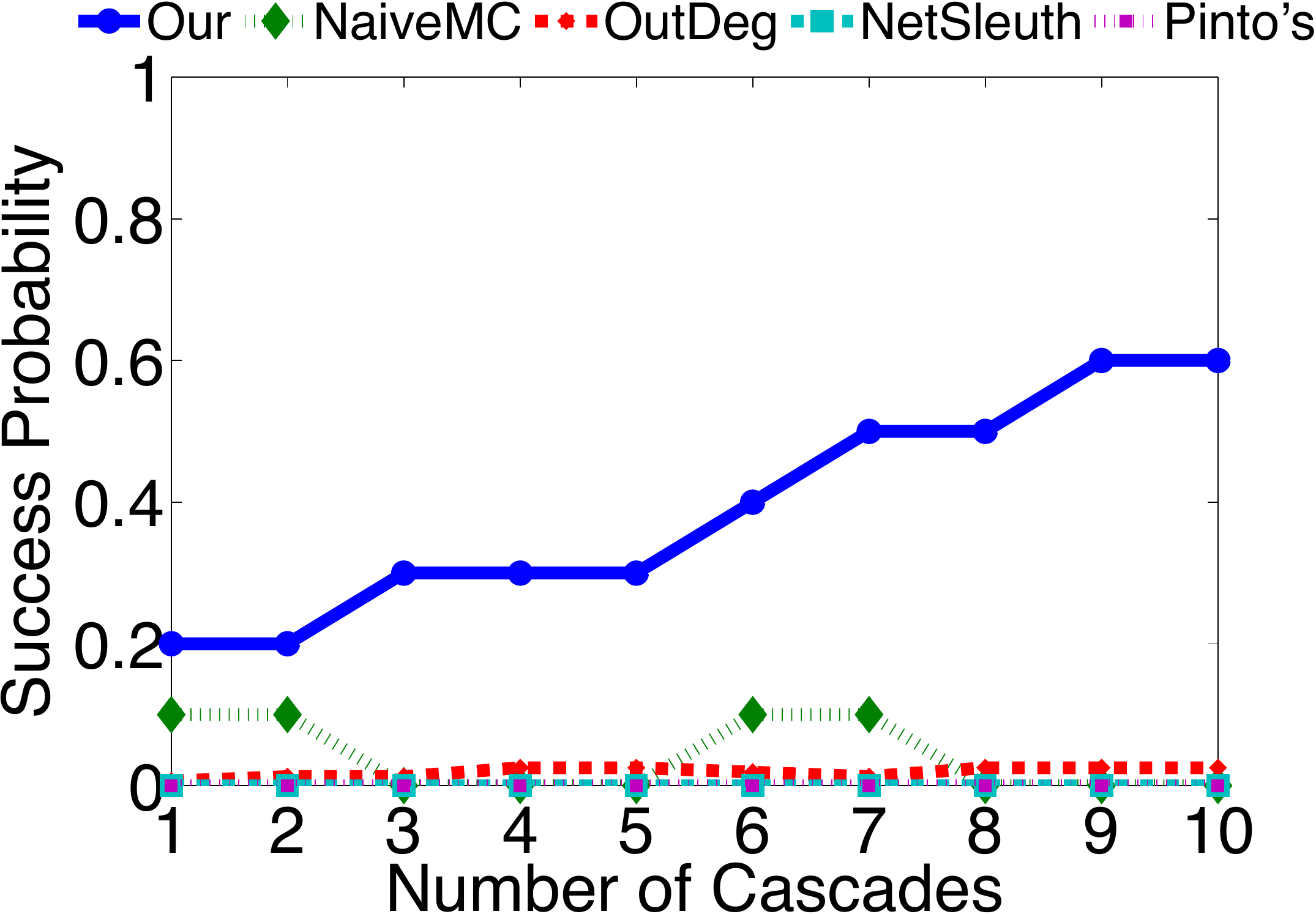} & 
    \includegraphics[width=0.245\textwidth]{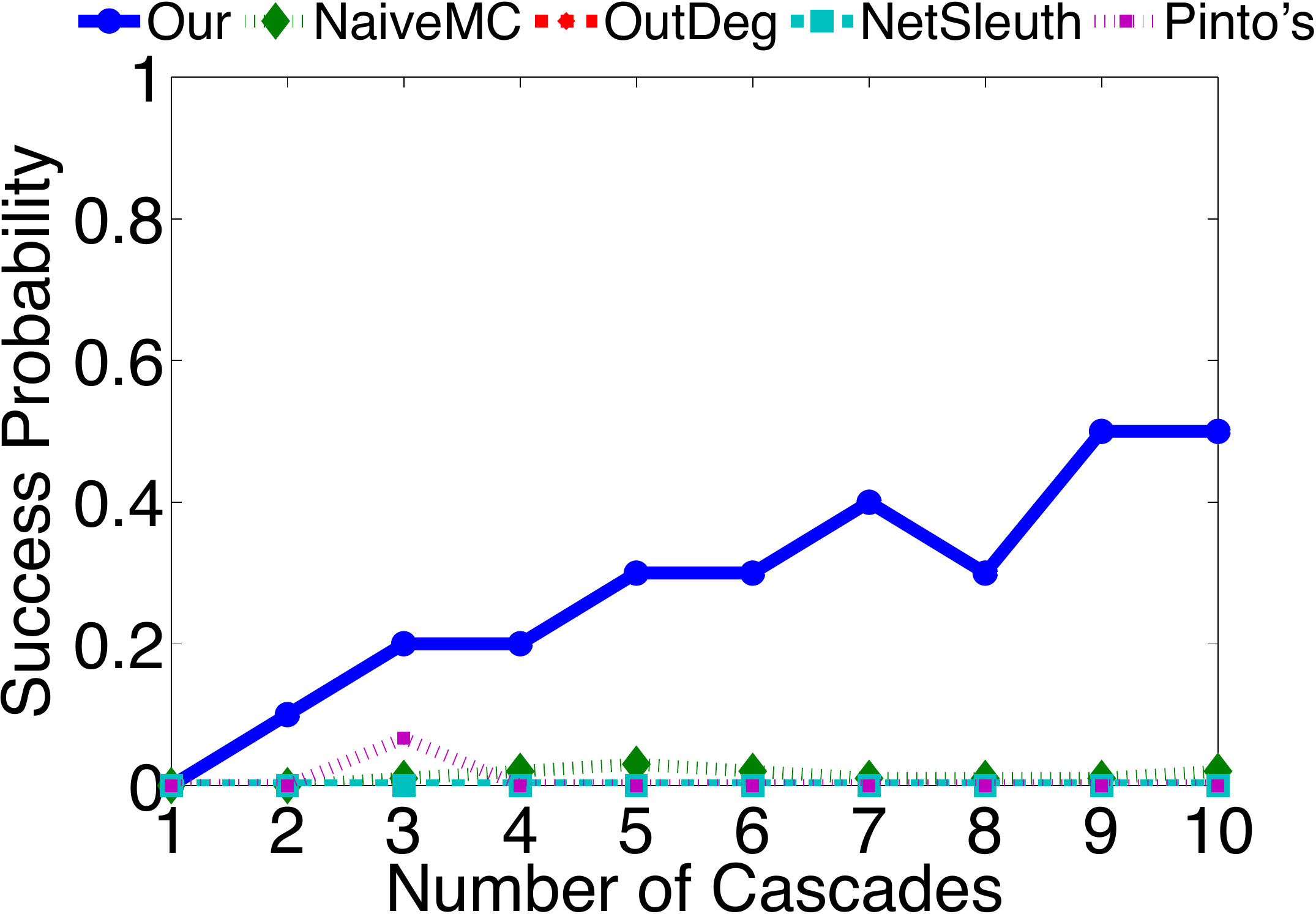} & 
    \includegraphics[width=0.245\textwidth]{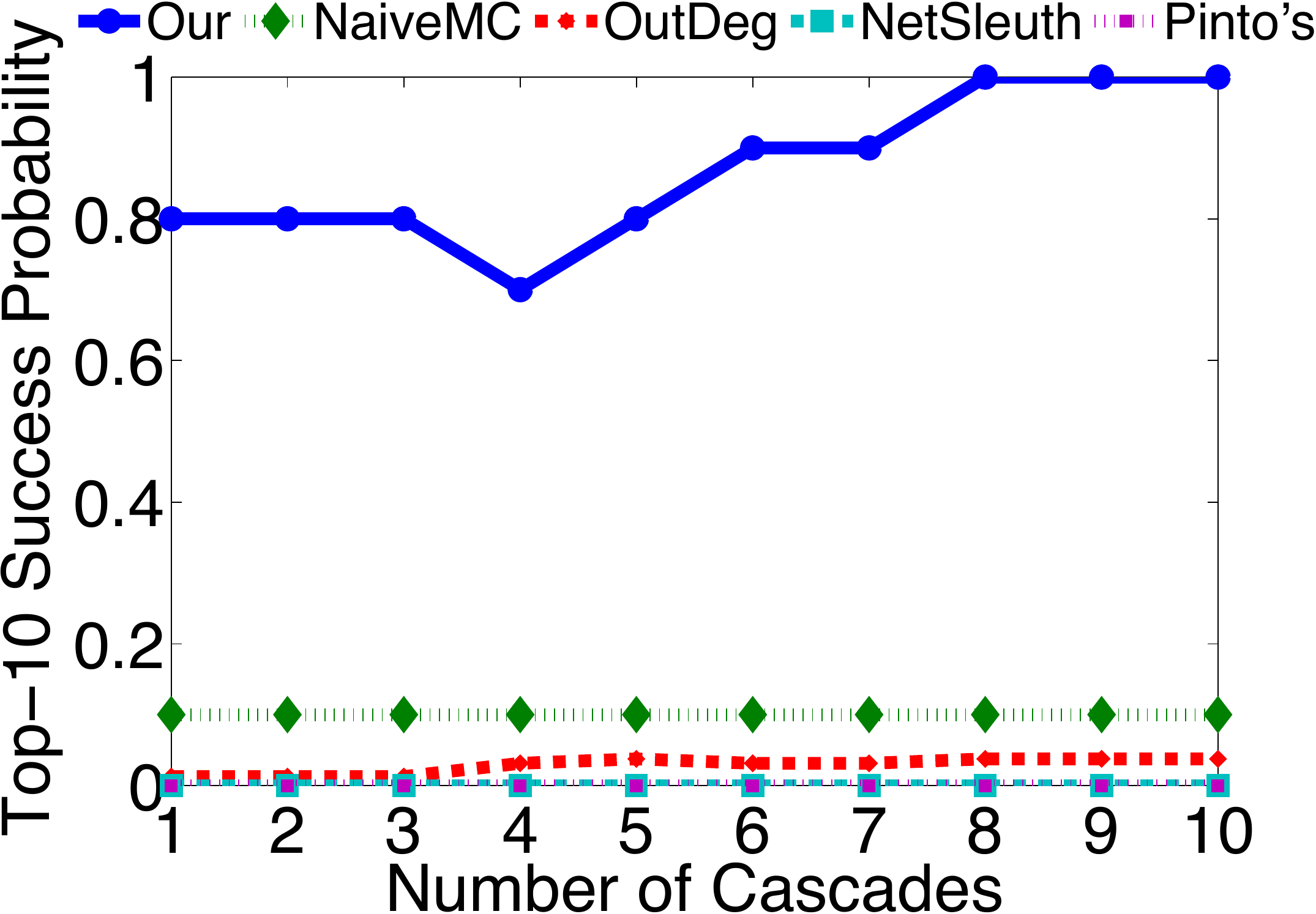} & 
    \includegraphics[width=0.245\textwidth]{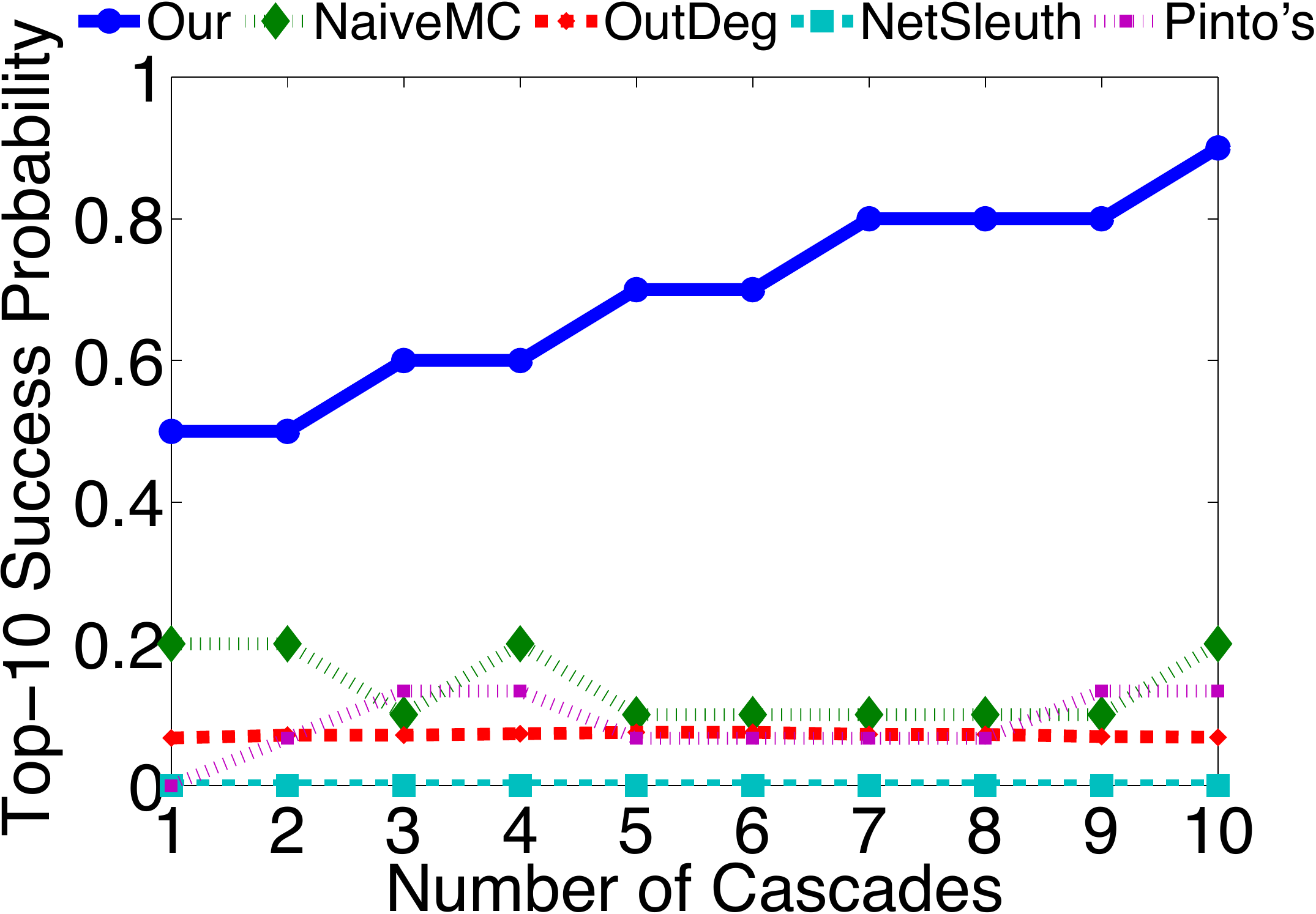} \\
    (a) SP, Random & (b) SP, Core-Periphery & (c) Top-10 SP, Random & (d) Top-10 SP, Core-Periphery
  \end{tabular}
  \caption{Success Probability (SP) and Top-10 Success Probability (Top-10 SP) for two types of Kronecker networks.} \label{fig:acc-cascades}
\end{figure*}

{\bf A Toy Example.}
We first consider a small 64-node hierarchical Kronecker network and vi\-sua\-lize the approxi\-mate likelihood given by Eq.~\ref{eq:sampling-simplified} against the number of 
observed cascades for each node in the network. We use $150$ Monte Carlo samples.
Figure~\ref{fig:grid} summarizes the results, where each square represents a node, the true source is marked with a star and the heat map represents normalized likelihoods in $[0, 1]$.
%
%
In this toy example, a single cascade is insufficient to detect the true source, since it has a relatively low likelihood. 
However, once more cascades are observed, the likelihood of the true source increases and ultimately become higher than all other nodes for $8$ cascades.
%

%
{\bf Accuracy.} 
Next, we evaluate the accuracy of our method in comparison with two state of the art me\-thods, \netsleuth~\citep{prakash2012} and Pinto'{}s method~\citep{pinto2012locating}, and 
two baselines in larger synthetic networks. 
The first baseline runs Montecarlo from each potential source and ranks them by counting the average maximum number of observed infected nodes that get infected in a time window 
equal to the length of observation window. Then, it ranks the potential sources according to the average value of this quantity, where the node with the highest value is the top
node.
The second baseline first finds all potential sources that can reach all observed infected nodes and then ranks them by decreasing out-degree, where the node with the highest
out-degree is the top node. 
%
%
\netsleuth~assumes the same infection probability $\beta$ over all the edges, which we set to $0.1$, following~\cite{prakash2012}. 
Pinto'{}s method similarly assumes that all pairwise transmission times come from the same Gaussian distribution and they require its mean to be much larger than its standard deviation in order to 
guarantee nonnegative transmission times. 
In their work, they set $\mu / \sigma = 4$, where $\mu$ and $\sigma$ are the mean and standard deviation, respectively. Since our fitted diffusion network contains edges with different 
transmission rates and thus different expected transmission time, we set the parameter $\mu$ to be the minimum expected value over all the edges. 
%

We used two measures of accuracy: success pro\-ba\-bi\-li\-ty and top-10 success probability. 
We define success probability as $P(\hat{s} = s^{*})$ and top-10 success probability as the probability that the true source $s^{*}$ is among the top-10 in terms of maximum likelihood or
ranking. For each network type, we estimated both measures by running our method on $10$ different random source sets.
Since \netsleuth and Pinto's method can only accept one observed cascade at a time, we run the methods independently for each individual cascade and then compute the top-1 and top-10 success 
probability based on all outputs for all cascades.
Figure~\ref{fig:acc-cascades} summarizes the results for two types of Kronecker networks against the number of observed cascades. Our method outperforms dramatically all others, achieving a success probability as high as $0.6$ and top-10 success probability of almost $1$. 
The low performance that state-of-the-art methods exhibit, in comparison with the validation within the corresponding papers, may be explained as follows: in both cases, the authors validated their algorithms with synthetic and real networks with large diameters, without long-range connections, such as 2-D grids~\citep{prakash2012} and spatial (geographical) networks~\citep{pinto2012locating}, where the source identification problem is much easier.
\begin{figure}[t]
        \centering
        \subfigure[Random]{\includegraphics[width=0.235\textwidth]{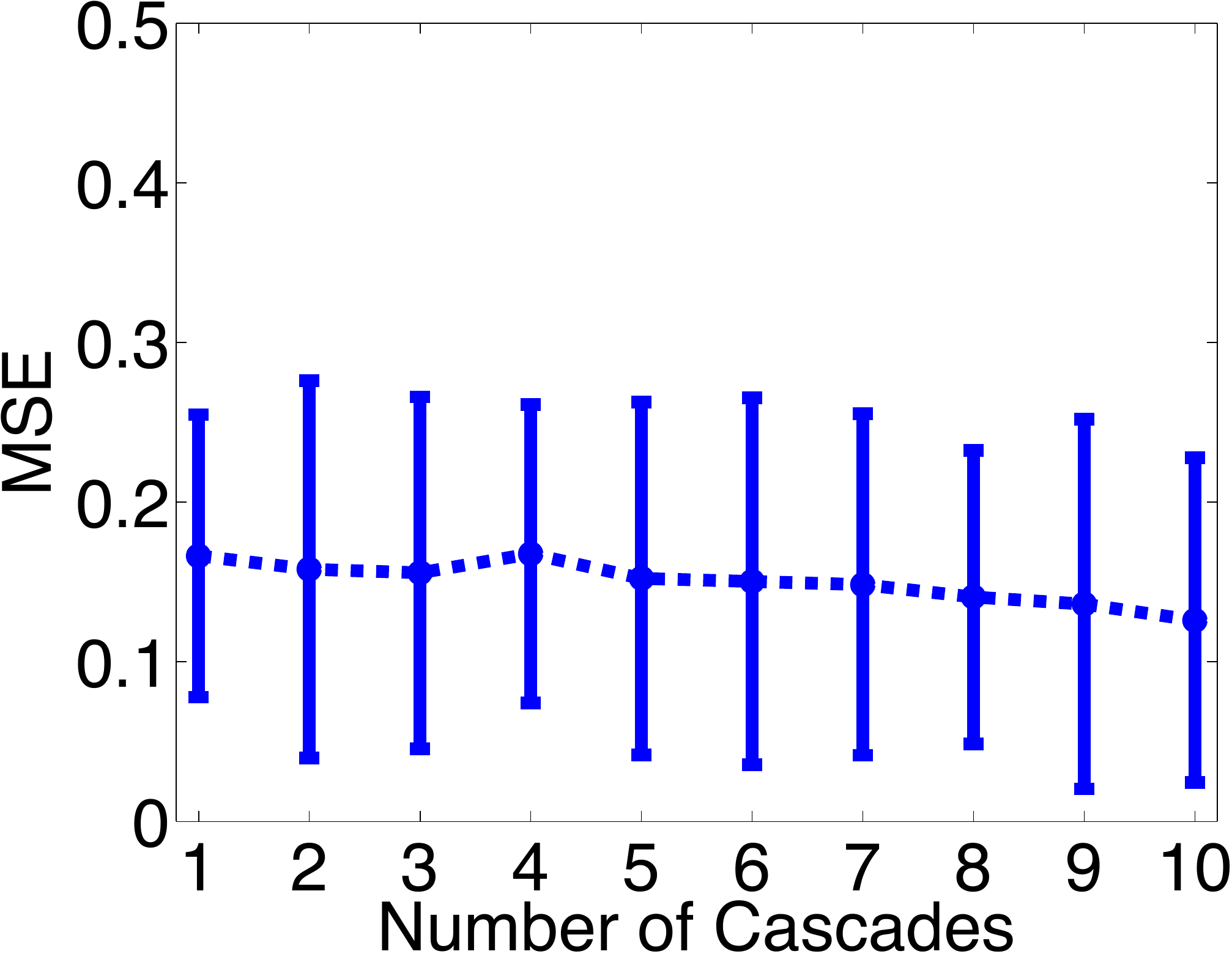}}
        \subfigure[Core-Periphery]{\includegraphics[width=0.235\textwidth]{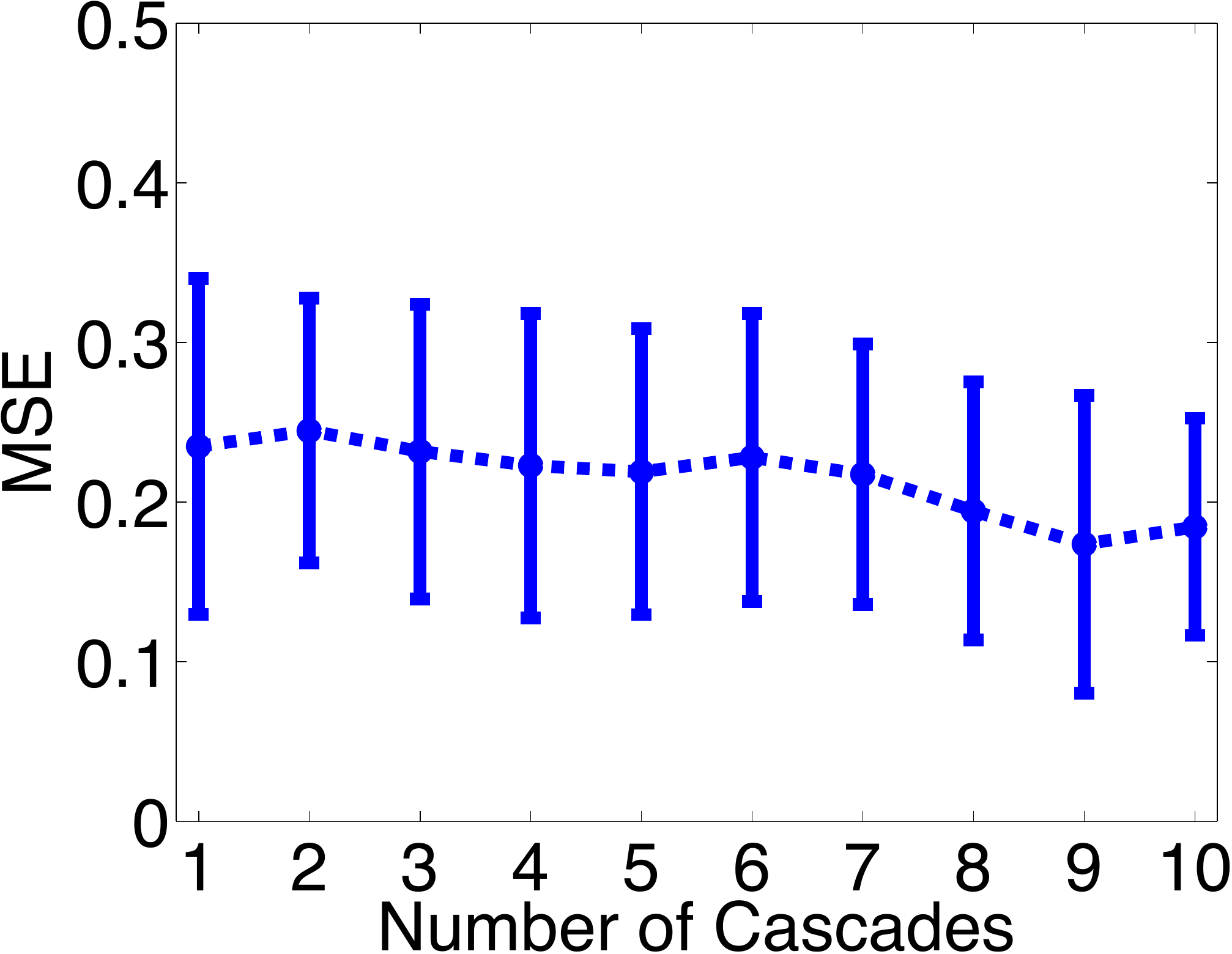}}
        \caption{Mean-squared error (MSE) on the estimation of $t_s$ for two types of Kronecker networks.} \label{fig:mse}
\end{figure}

\begin{figure*}[!t]
        \centering
        \subfigure[SP]{\includegraphics[width=0.28\textwidth]{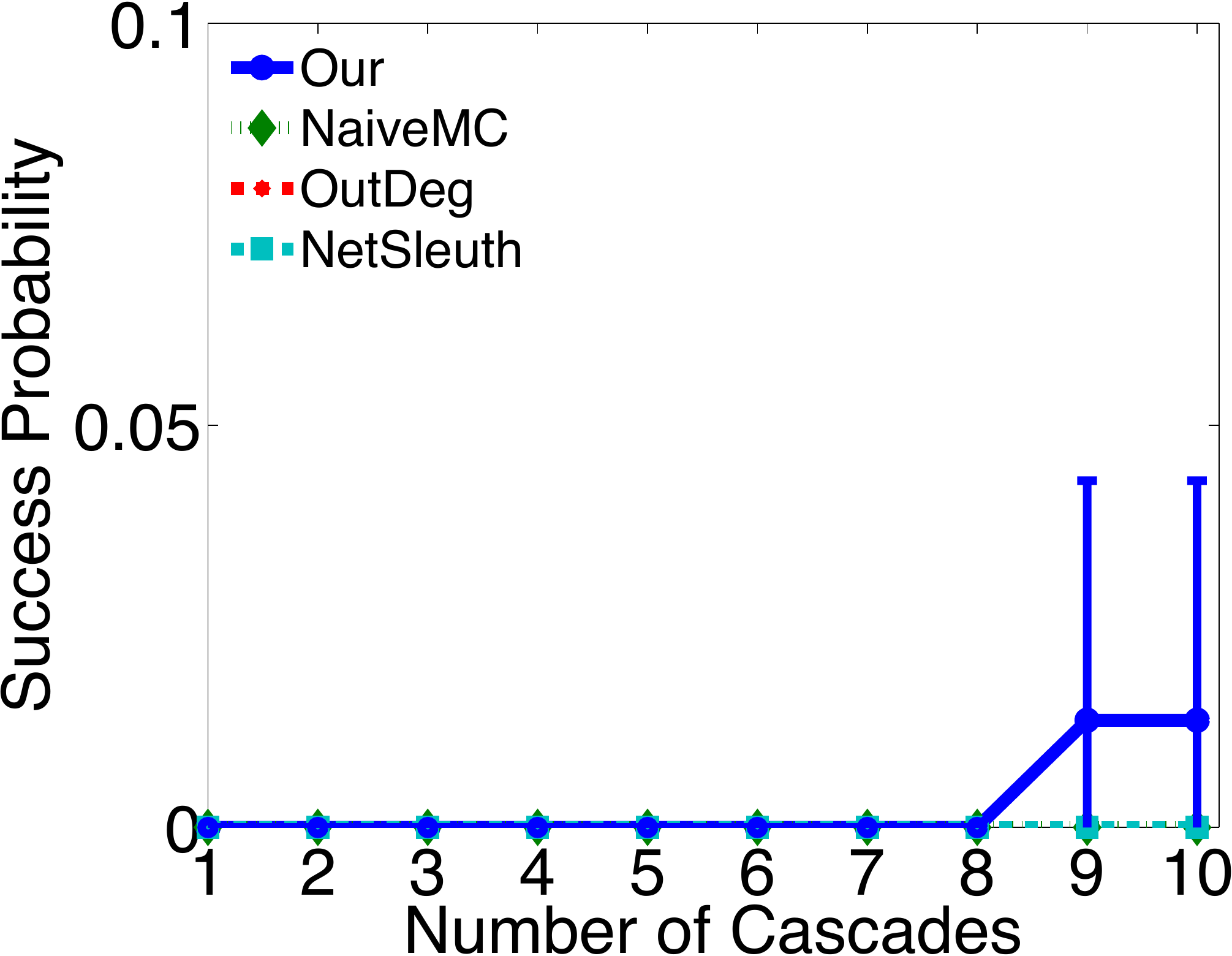}} \hspace{2mm}
        \subfigure[Top-10 SP]{\includegraphics[width=0.28\textwidth]{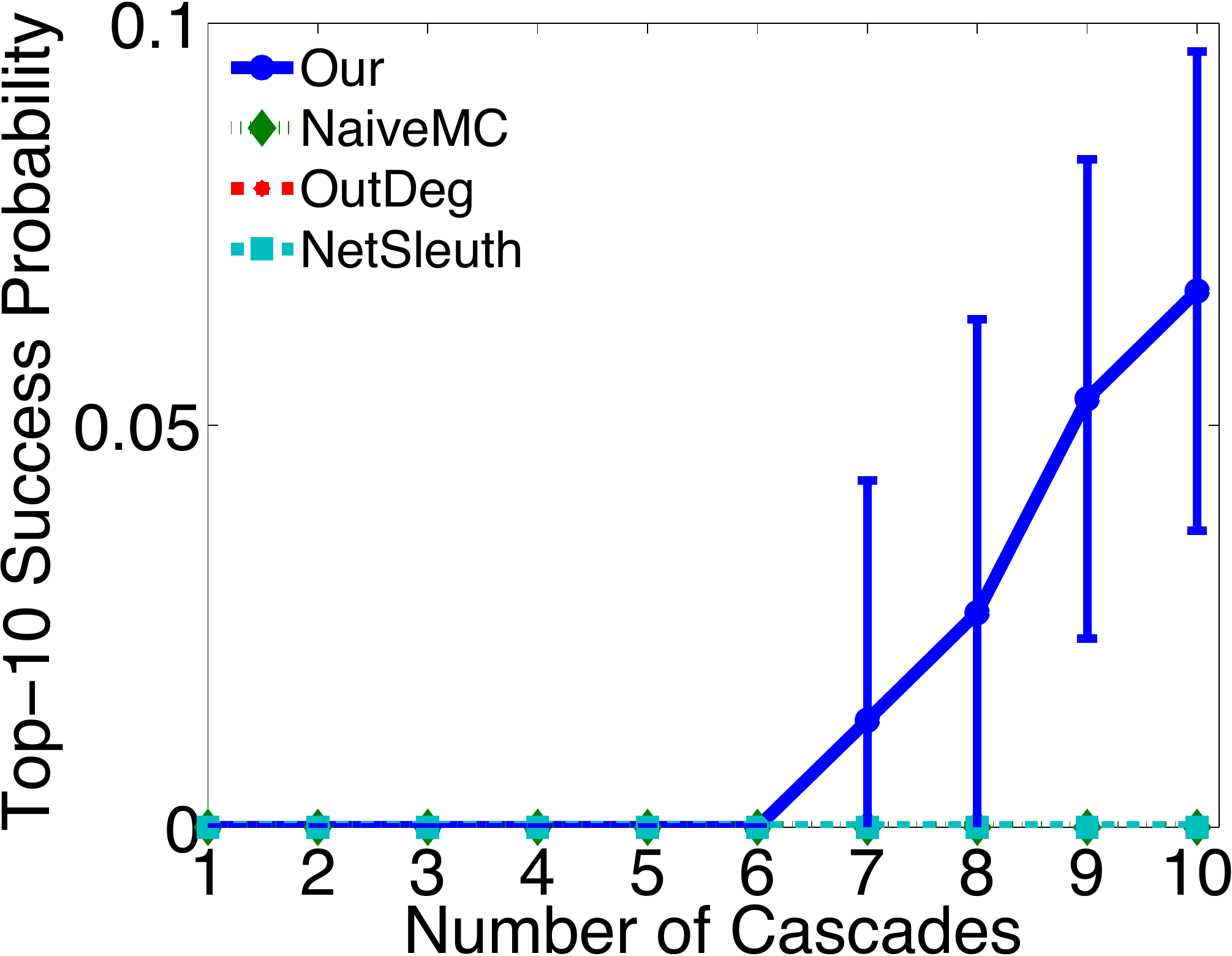}} \hspace{2mm}
        \subfigure[MSE]{\includegraphics[width=0.28\textwidth]{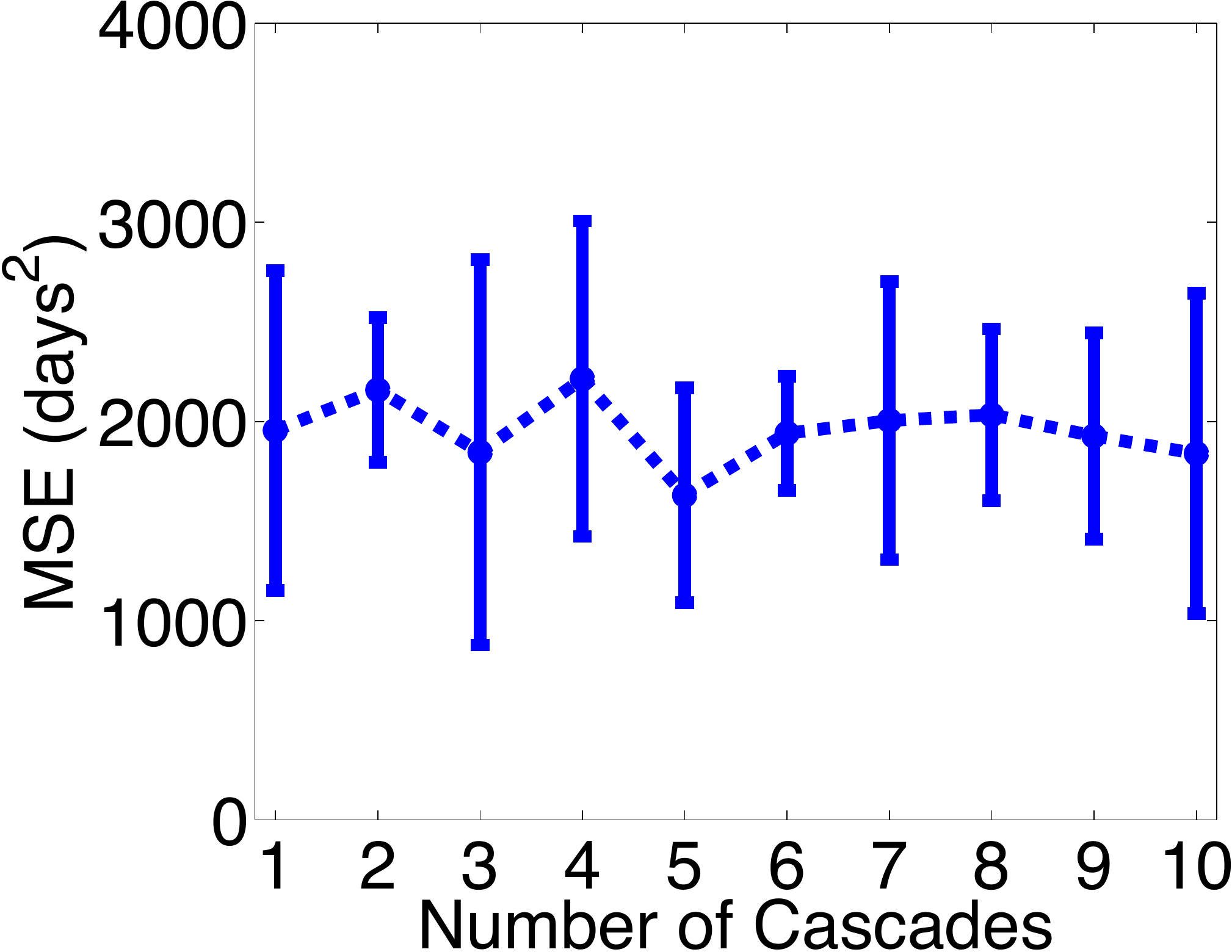}}
        \caption{Success Probability (SP), Top-10 Success Probability (Top-10 SP) and mean-squared error (MSE) on the estimation of $t_s$ for real cascade data.} \label{fig:real-all}
\end{figure*}

\xhdr{Source Infection Time Estimation} We also eva\-luate how accurately our method infers the infection time of the true source by computing the mean square error (MSE),
$E_{s^{*}}\left[ (t_{s^{*}} - \hat{t}_{s^{*}})^2 \right]$, estimated by running our method on $10$ different random sources. Here, we do not compare with other
competitive methods since they do not provide an estimate of the infection time of the true source.
Figure~\ref{fig:mse} shows the MSE of the estimated infection times of the true source for the same networks as above against the number of cascades.

\subsection{Experiments on Real Data}
%
{\bf Experimental Setup.} We focus on the spread of memes, which are a short textual phrases (like, ``lipstick on a pig'') that travel almost intact through the Web~\citep{leskovec2009kdd}.
We experiment with a large meme dataset\footnote{Data is available at \url{http://snap.stanford.edu/infopath/}}, which traces the spread of memes 
across 1,700 popular mainstream media sites and blogs~\citep{manuel13wsdm}.
The dataset classifies memes per topic, and associates each meme $m$ to an information cascade $\casc_m$, which is simply a record of times when sites first 
mentioned meme $m$.
We proceed as follows. We first infer an underlying diffusion network per topic using \netrate, a well-known network inference method~\citep{manuel11icml},
using all observed information cascades.
We then use these inferred networks along with a percentage of the infections of large cascades to infer the source of these cascades. We select $15$ sources, each of them 
having at least 10 long cascades. 
Here, by long cascade we mean possessing more than 27 nodes. 
The results are averaged over 5 runs, randomizing the selection of the observed nodes, we consider that 10\% of the infected nodes are observed and utilize $500$ samples 
to approximate the likelihood.

\xhdr{Accuracy} We evaluate the accuracy of our method in comparison with \netsleuth and the same baselines as in the synthetic experiments, using success probability and 
top-10 success probability.
Unfortunately, we cannot compare to Pinto'{}s method because it requires the identity of the true parent for each observed node in each cascade, and this is not available in real 
cascade data.
Figure~\ref{fig:real-all} summarizes the results.
Surprisingly, neither \netsleuth nor the baselines succeed at detecting cascade sources in real data, even with $10$ observed cascades; they output solutions with an (almost) 
zero (top-10) success probabilities. 
In contrast, our method achieves a non-zero (top-10) success probability as long as we observe more than $8$ and $6$ cascades respectively, a fairly low number of cascades in 
this scenario. Even then, the performance of our method in terms of success and top-10 success probability may seem low at first, however, we would like to highlight how difficult 
the problem we are trying to solve is, by considering the performance of two simple random guessers.
A first random guesser who chooses the source uniformly at random from all nodes in the network would succeed with probability $1/1700 = 5.8 \times 10^{-4}$, almost $20$ times 
less accurate than our method. 
A second random guesser that chooses the source uniformly at random among the nodes from whom the observed nodes are reachable would succeed with probability 
$1/425 = 2.4 \times 10^{-3}$, almost $5$ times less accurate. 
The same argument for top-10 success probability shows $12$ times improvement in accuracy compared to naive guesser and $3$ times improvement in comparison to the more 
clever one.
Finally, our method'{}s MSE values indicate that our method is able to find the source infection time within an accuracy of $\sqrt{2000} \approx 45$ days. We find this quite remarkable 
given that the cascades we considered typically unfold during a 1-year period.

%
%

\section{CONCLUSIONS}
\label{sec:conclusions}
We propose a two-stage framework for detecting the source of a cascade in continuous-time diffusion networks, which improves dramatically over previous state-of-the-arts in terms of detection 
accuracy. Our framework cast the problem as a maximum likelihood estimation problem and then find optimal solutions very efficiently using an importance sampling approximation to the objective 
and an optimization procedure that exploits the structure of the problem. 
Our work opens many interesting venues for future work. For example, it would be useful to extend our method to support cascades with multiple sources and other continuous-time models 
different than the continuous-time independent cascade model~\citep{manuel11icml}. Also, a theoretical analysis of our importance sampling scheme is also interesting. Finally, it would be interesting 
to apply the current framework to other real-world datasets.

\subsection*{Acknowledgements}
This work was supported in part by NSF/NIH BIGDATA 1R01GM108341, NSF IIS-1116886, NSF CAREER IIS-
1350983 and a Raytheon Faculty Fellowship to L.S.

\bibliographystyle{abbrvnat}
\bibliography{refs}

\clearpage
\newpage
\onecolumn

\begin{appendix}
\label{sec:appendix}
\section{Proof of Lemma~\ref{lemma1}} \label{app:lemma1-proof}
Suppose there are two stationary points, \ie, $\phi_L'(x) = \phi_L'(y) = 0$, thus, by continuity of $\phi_L(.)$ in  $( t_{s_i}, t_{s_{i+1}})$ there must be a $z \in (x, y)$ such that $\phi_L''(z) = 0$. We 
show it is a contradiction as
\begin{equation}
\phi_L''(t_s) = \sum_l^L \gamma_l \beta_l^2 e^{\beta_l t_s} > 0
\end{equation}
for all $1 \le l \le L$.

\section{Proof of Lemma~\ref{lemma2}} \label{app:lemma2-proof}
Assume we would like to find the maximizer of $\phi_L(.)$ in interval $(a,b)$ and consider two points at one-third and two-third of the interval, \ie, $c = a + \frac{b-a}{3}$ and $d = a + 2\frac{b-a}{3}$. 
It can be easily shown that, if $\phi_L(c) < \phi_L(d)$, then the maximizer will be on interval $(c, b)$ and, if $\phi_L(c) < \phi_L(d)$, then the maximizer must lie on interval  $(a,d)$. 
Therefore, by two evaluations, we can shrink the interval containing the maximizer by a factor of $\frac{2}{3}$. 
Then, to reach the $\epsilon$-neighborhood of the real maximizer, we need evaluate the function $2*r$ times, where 
\begin{equation} 
(t_{s_{i+1}}- t_{s_{i}})(2/3)^{r} < \epsilon.
\end{equation}
This will prove our claim.

\section{Additional Experimental Results} \label{app:experiments}
In this section, we provide additional experimental results on synthetic data, including an evaluation of 
the performance of our method against the percentage of observed infections and the number of Montecarlo
samples, as well as a scalability analysis.

\begin{figure*}[h]
        \centering
        \subfigure[Random]{\includegraphics[width=0.27\textwidth]{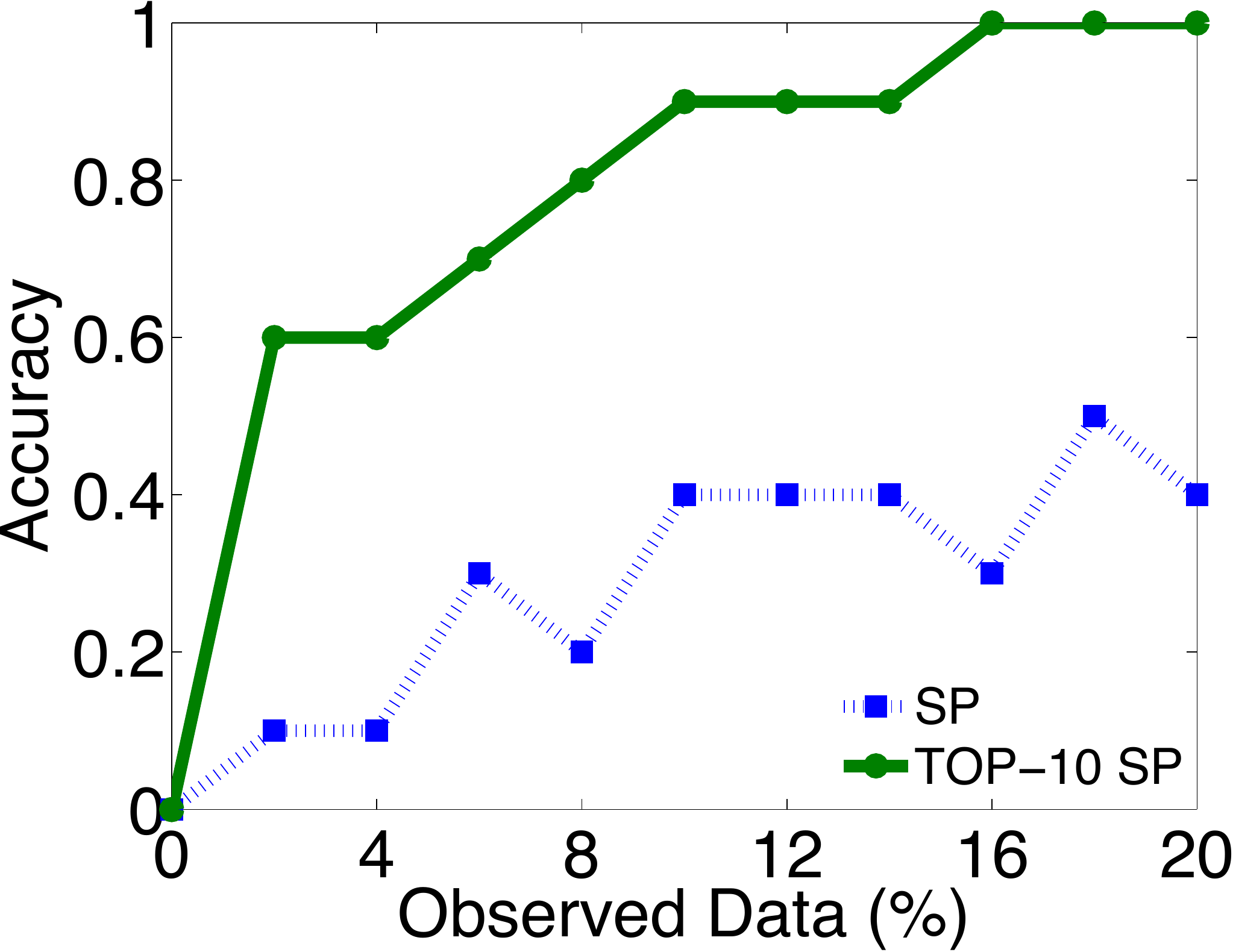}}
        \subfigure[Hierarchical]{\includegraphics[width=0.27\textwidth]{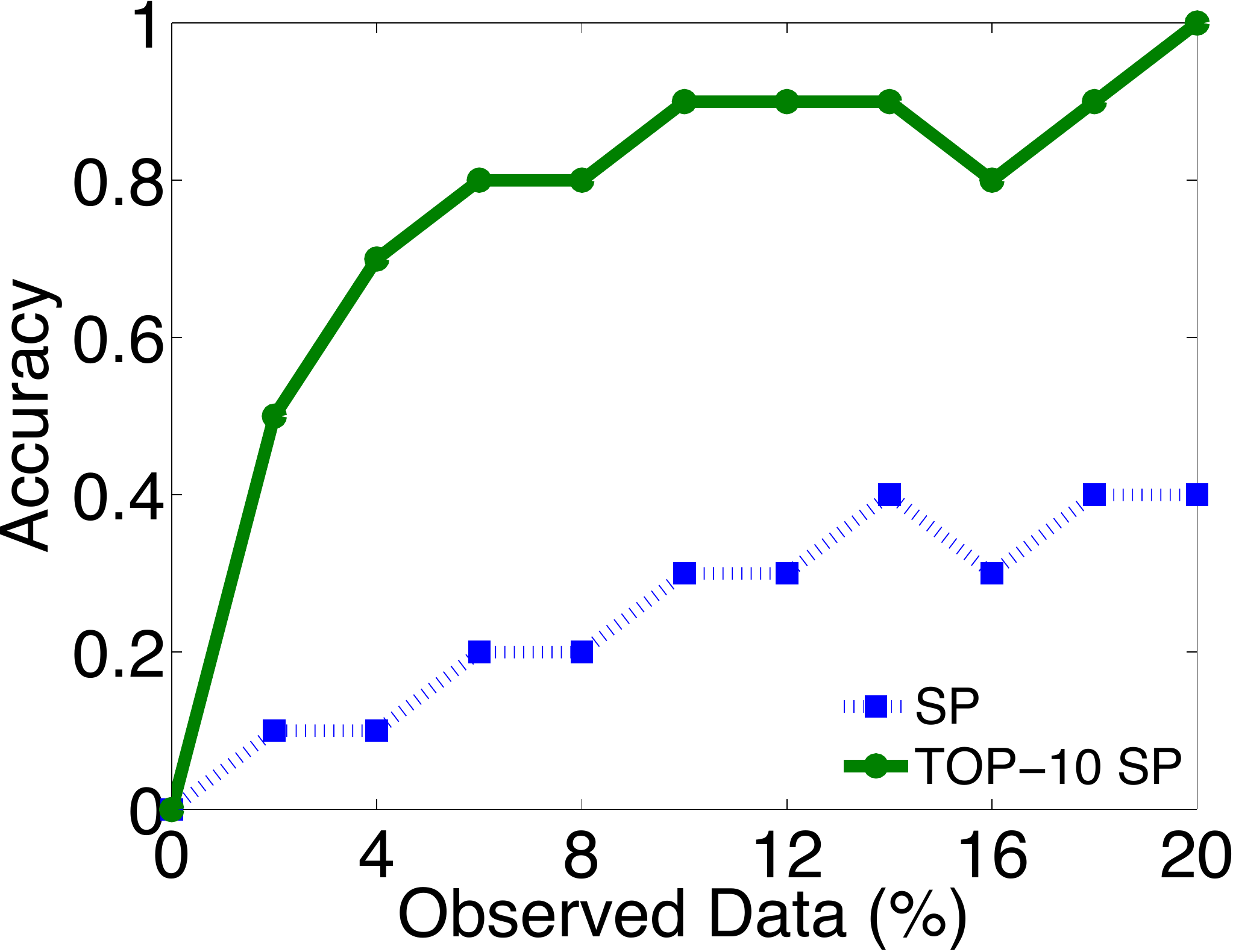}}
        \subfigure[Core-periphery]{\includegraphics[width=0.27\textwidth]{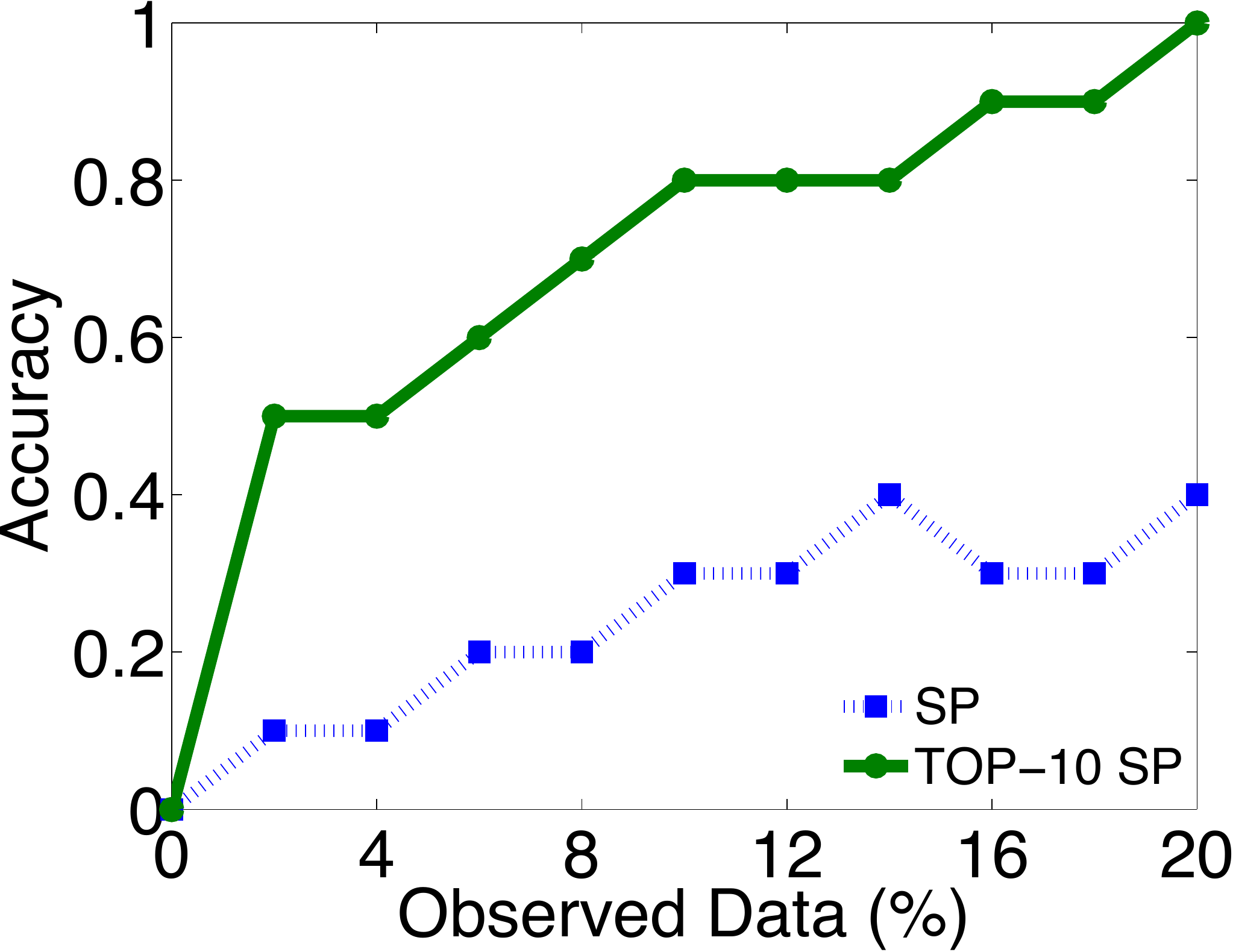}}
        \caption{Accuracy vs. \% observed infections.} \label{fig:acc-vs-observed}
\end{figure*}

\xhdr{Performance vs. percentage of observed infections}
Intuitively, the greater the number of observed infections, the more accurately our method can infer the true source and its infection time.
Figure~\ref{fig:acc-vs-observed} confirms this intuition by showing the success probability against percentage of observed infections.
However, we also find that the greater is the percentage of observed infections, the smaller is the effect of observing additional infections; a diminishing return property. 

\begin{figure*}[h]
        \centering
        \subfigure[Random]{\includegraphics[width=0.27\textwidth]{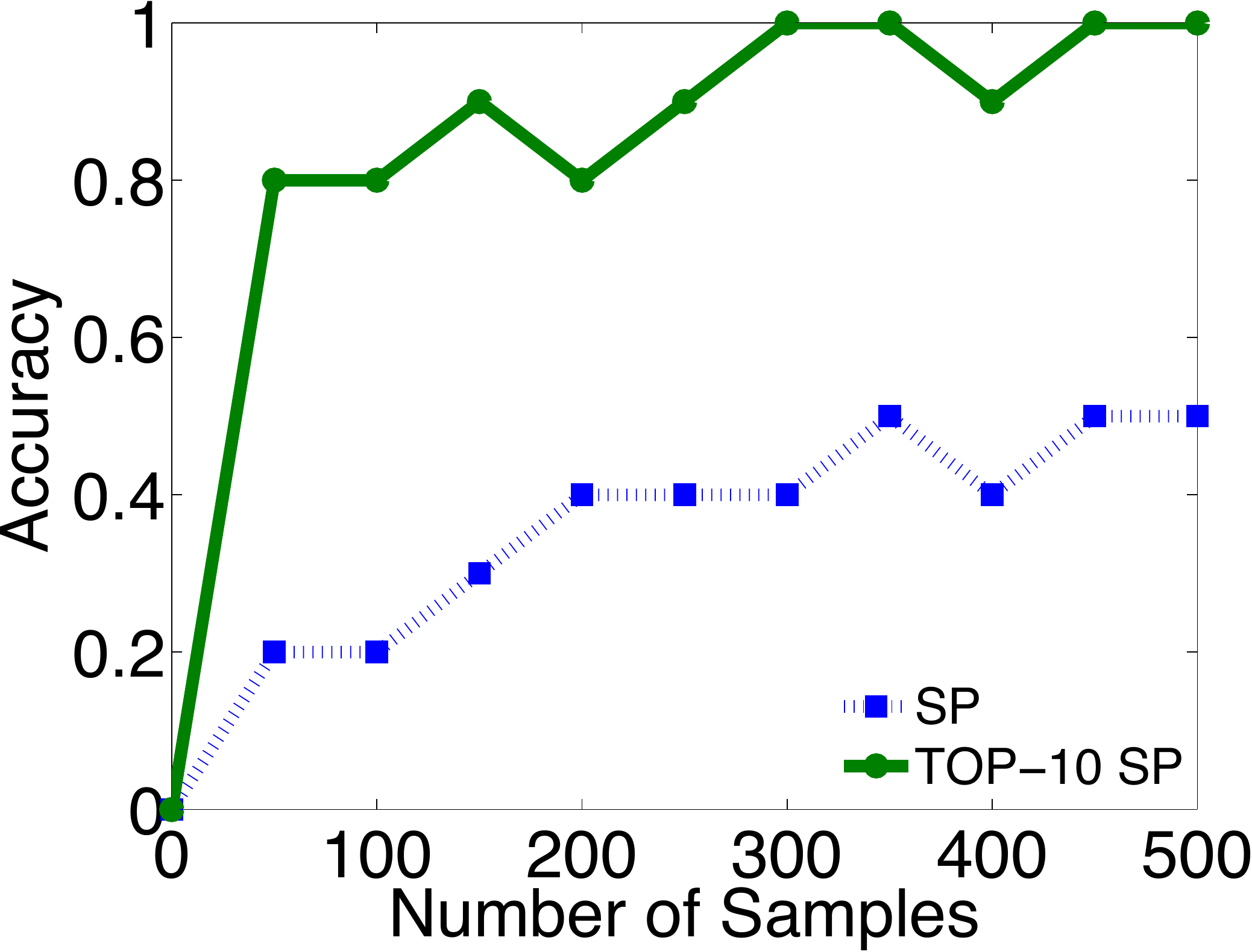}}
        \subfigure[Hierarchical]{\includegraphics[width=0.27\textwidth]{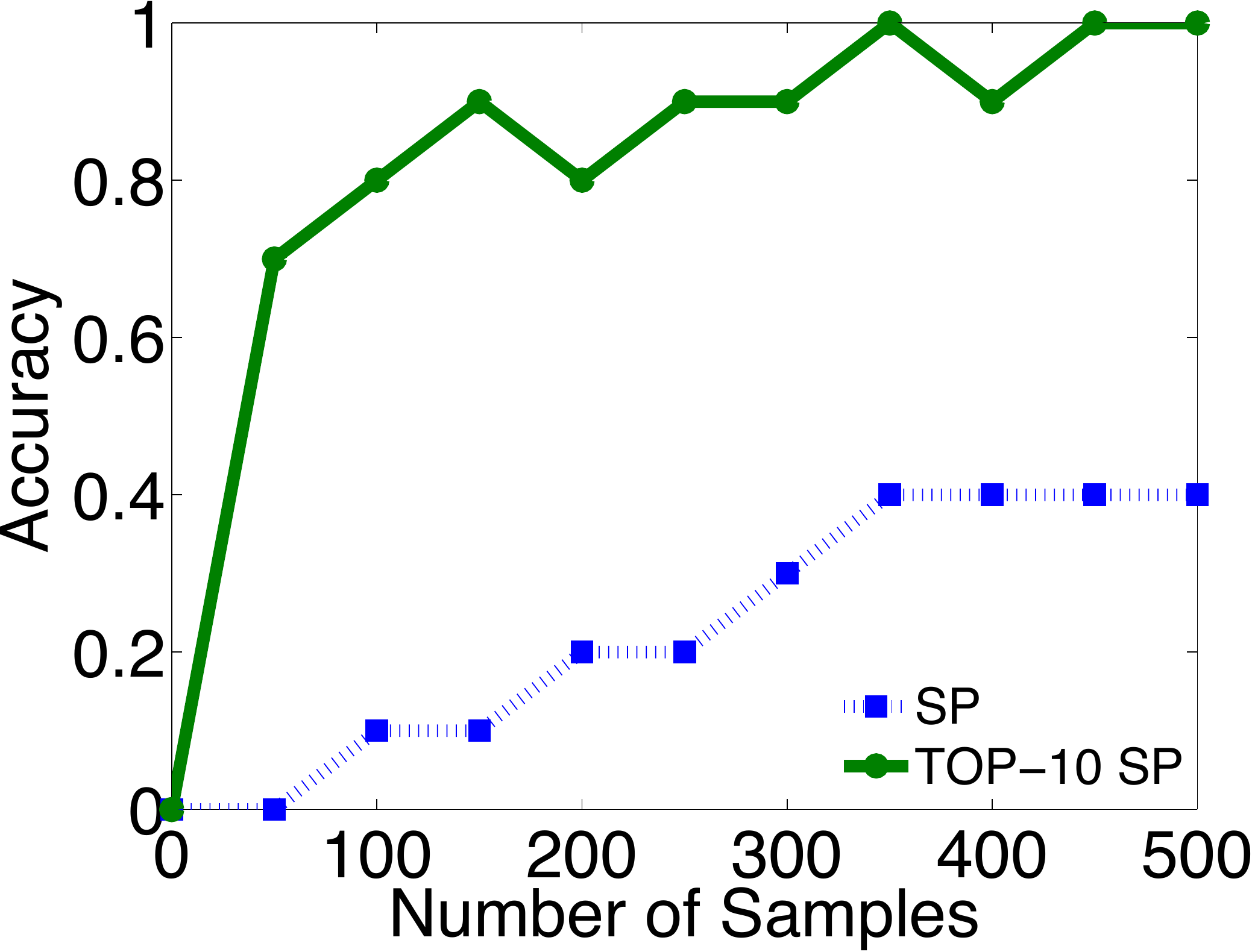}}
        \subfigure[Core-Periphery]{\includegraphics[width=0.27\textwidth]{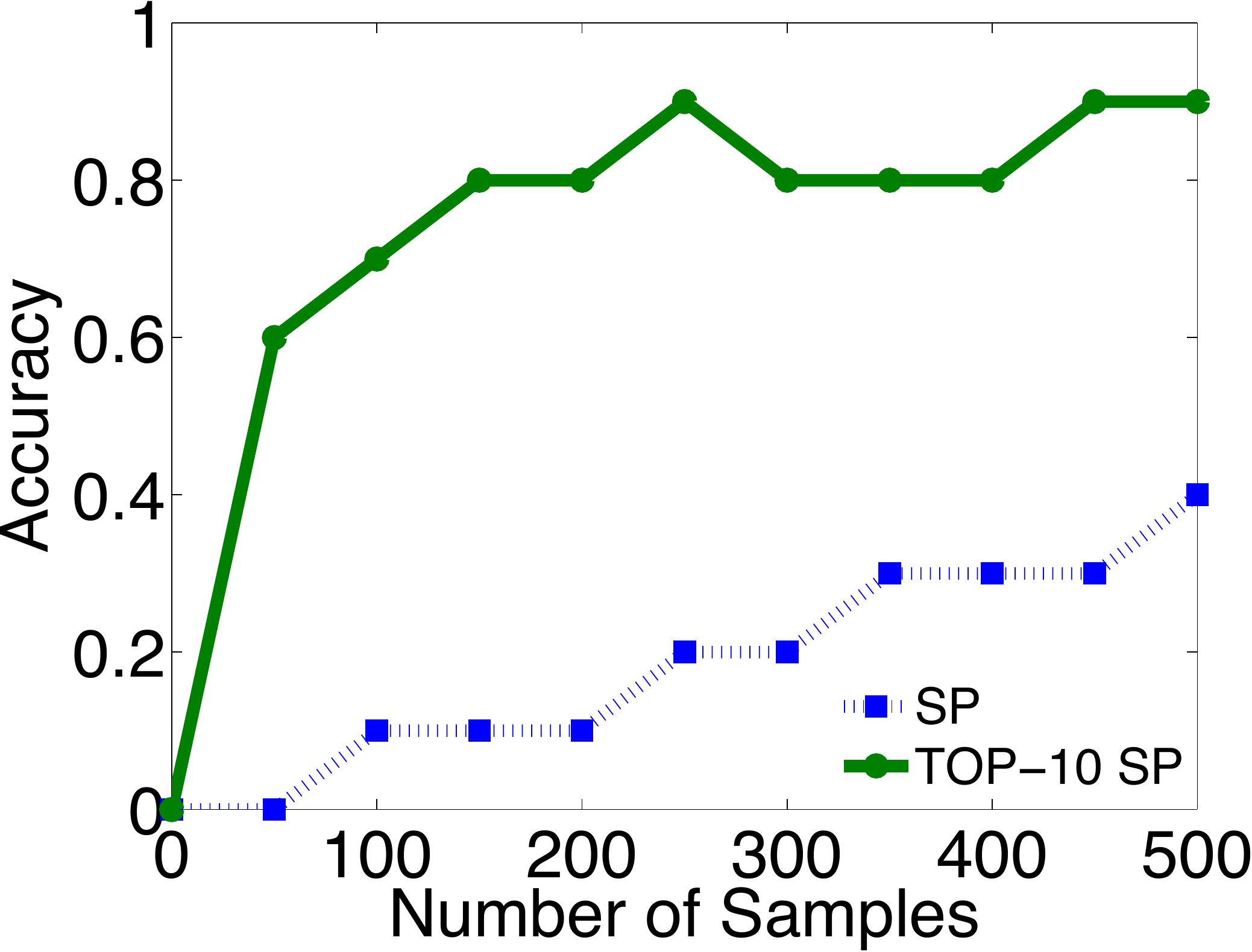}}
        \caption{Accuracy vs. number of samples.} \label{fig:acc-vs-samples}
\end{figure*}

%
\xhdr{Performance vs. number of Montecarlo samples}
Drawing more transmission time samples $\{ \tau_{ji} \}_{(j, i) \in \Ecal}$ leads to a better estimate of Eq.~\ref{eq:sampling}, and thus a greater 
accuracy of our method.
Figure~\ref{fig:acc-vs-samples} shows the success probability against number of samples.
Importantly, we observe that as long as the number of samples is large enough, the performance of our method quickly flattens and does not depend on
the number of samples any more.




%

\begin{figure*}[h]
        \centering
        \subfigure[Random]{\includegraphics[width=0.27\textwidth]{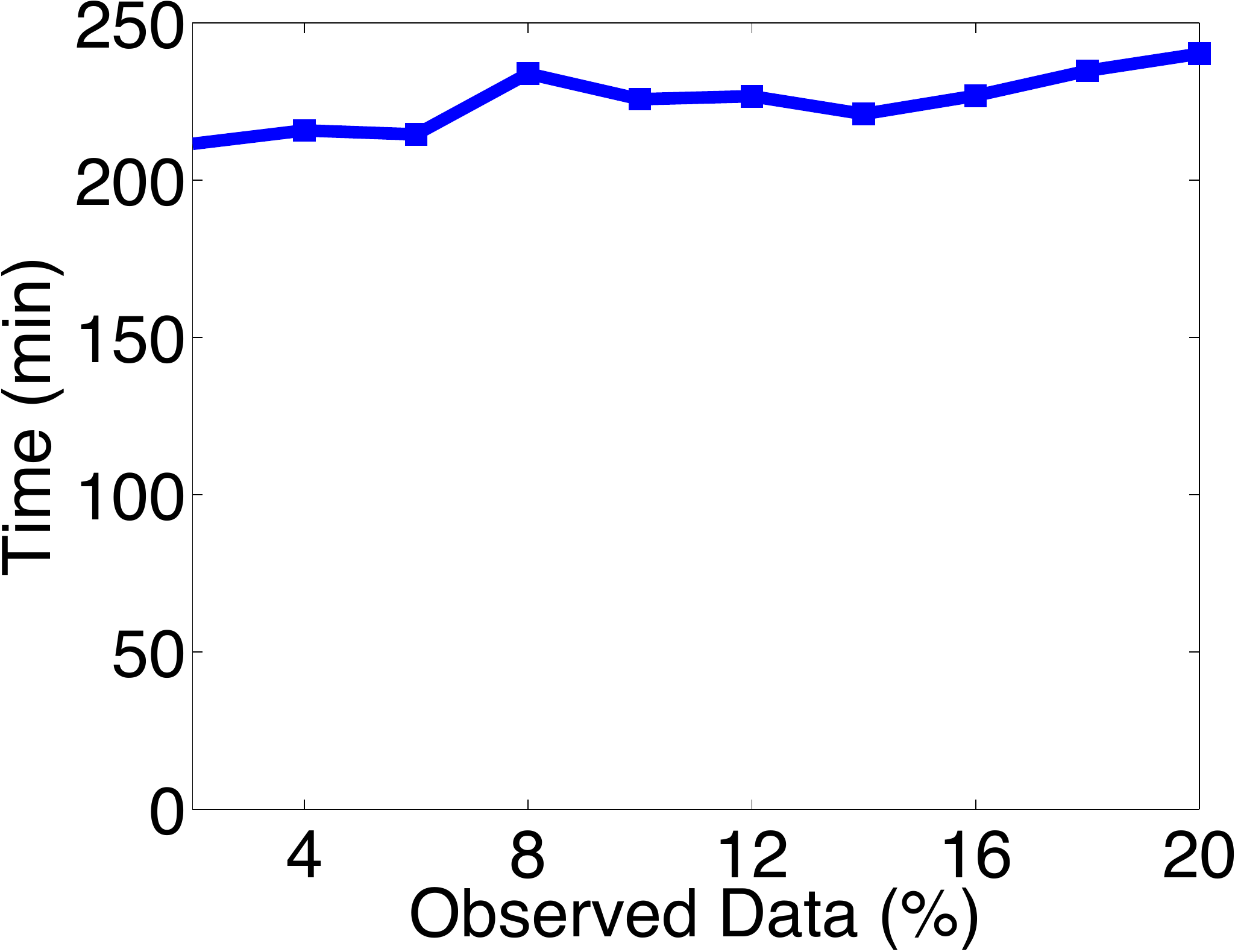}}
        \subfigure[Hierarchical]{\includegraphics[width=0.27\textwidth]{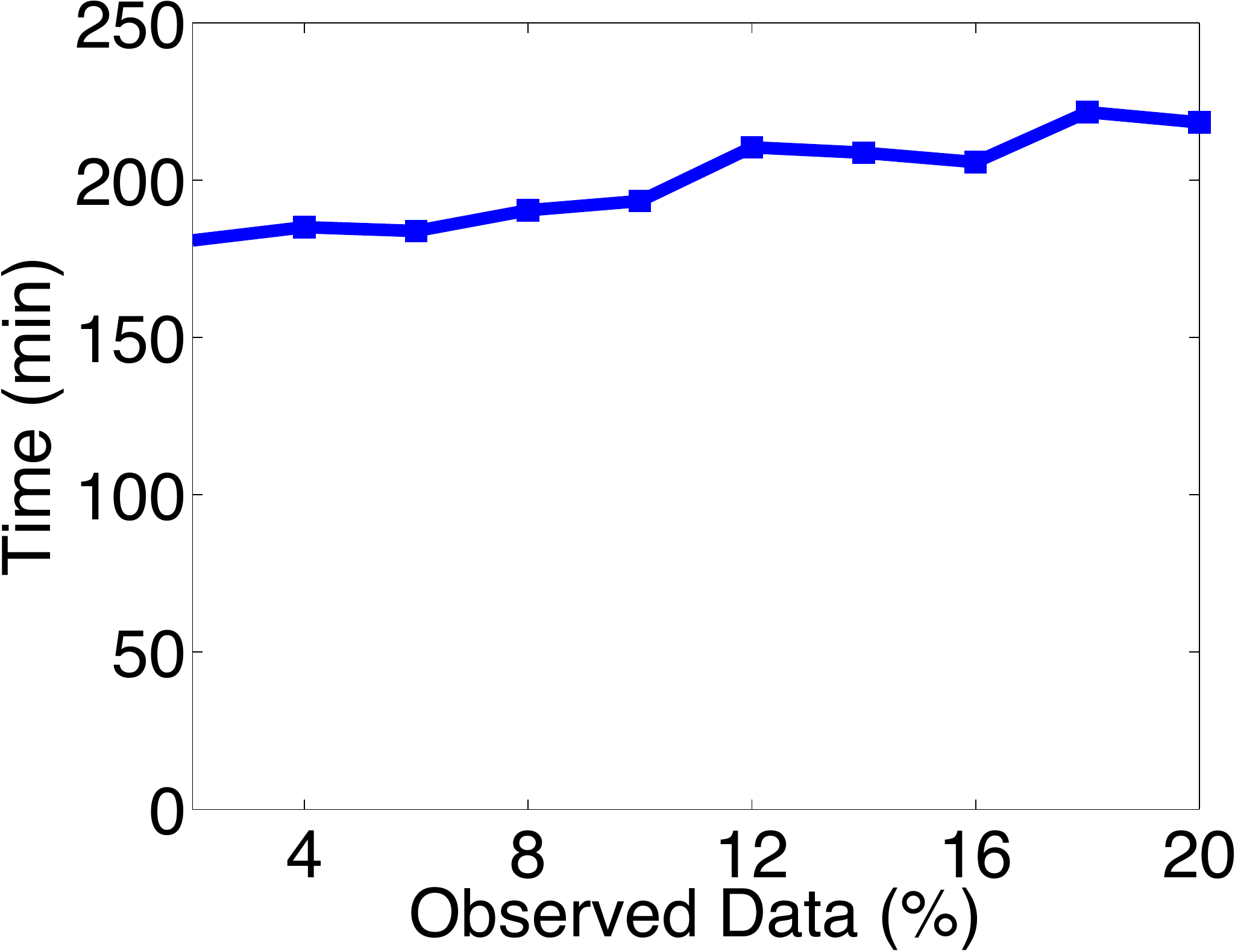}}
        \subfigure[Core-periphery]{\includegraphics[width=0.27\textwidth]{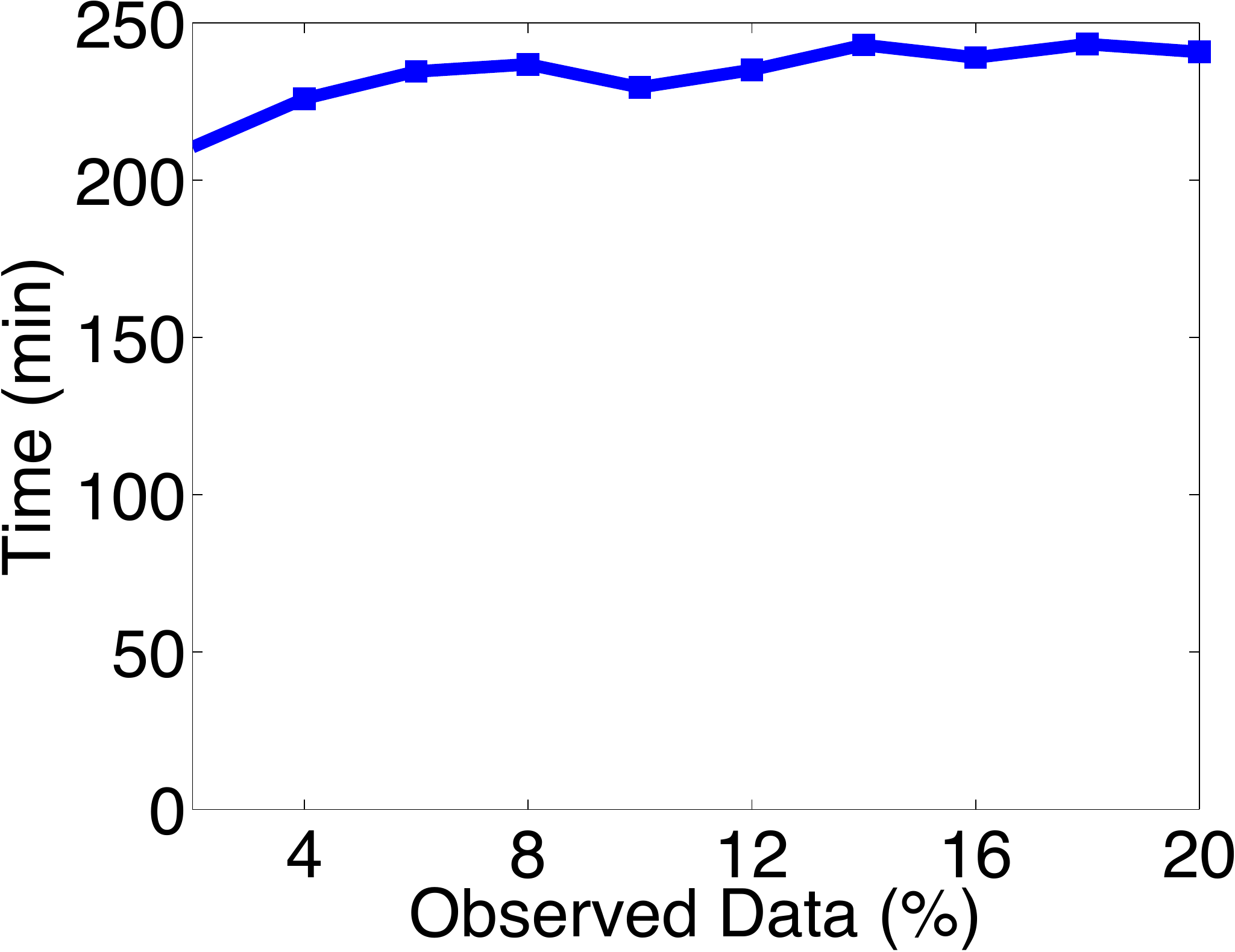}}
        \caption{Running time vs. \% observed infections.} \label{fig:time-vs-observed}
\end{figure*}

\xhdr{Running time vs. percentage of observed infections} 
Figure~\ref{fig:time-vs-observed} plots the average running time to infer the source of a single cascade against the percentage of observed
infections. Perhaps surprisingly, the running time barely increases with the percentage of observed infections.

\begin{figure*}[h]
        \centering
        \subfigure[Random]{\includegraphics[width=0.27\textwidth]{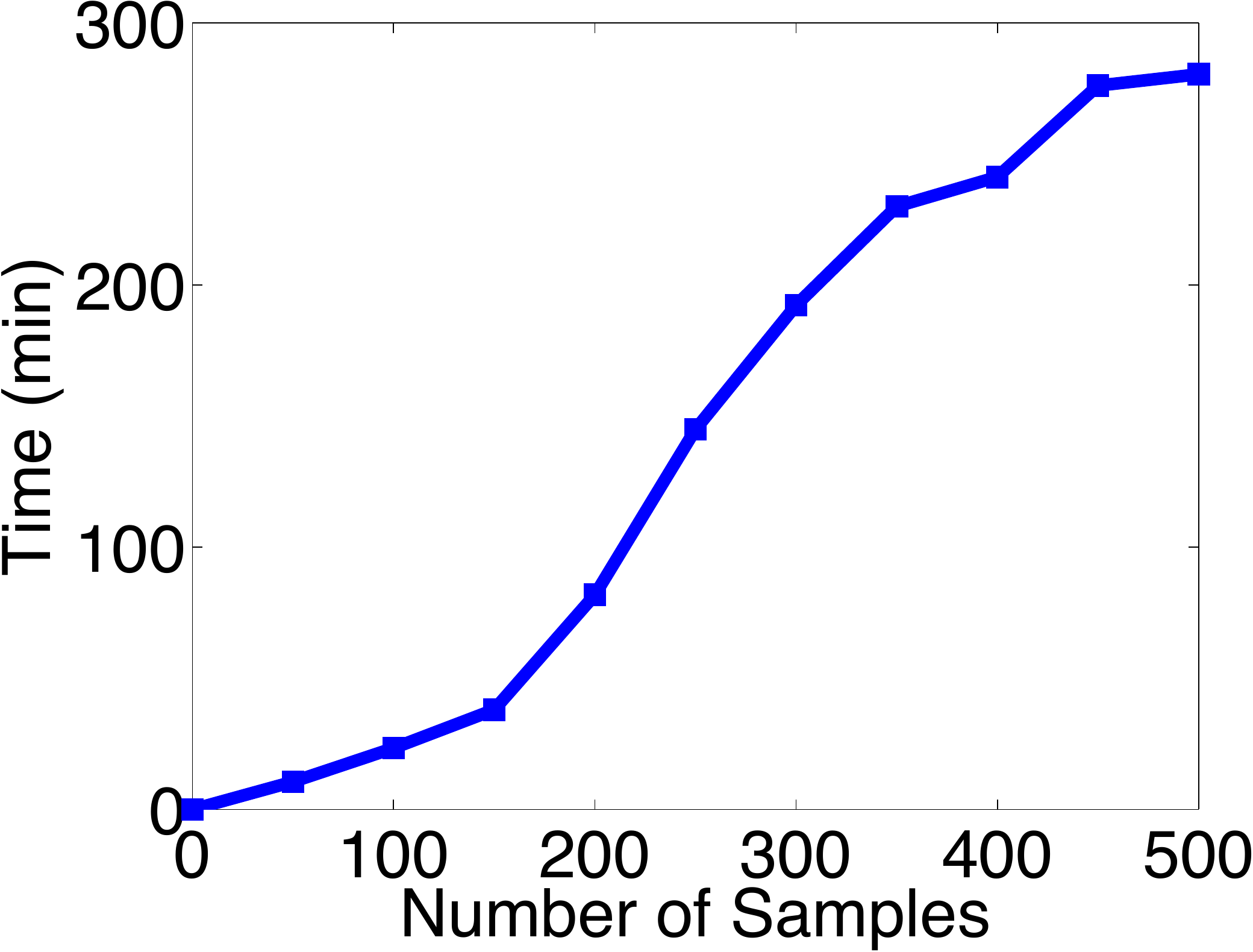}}
        \subfigure[Hierarchical]{\includegraphics[width=0.27\textwidth]{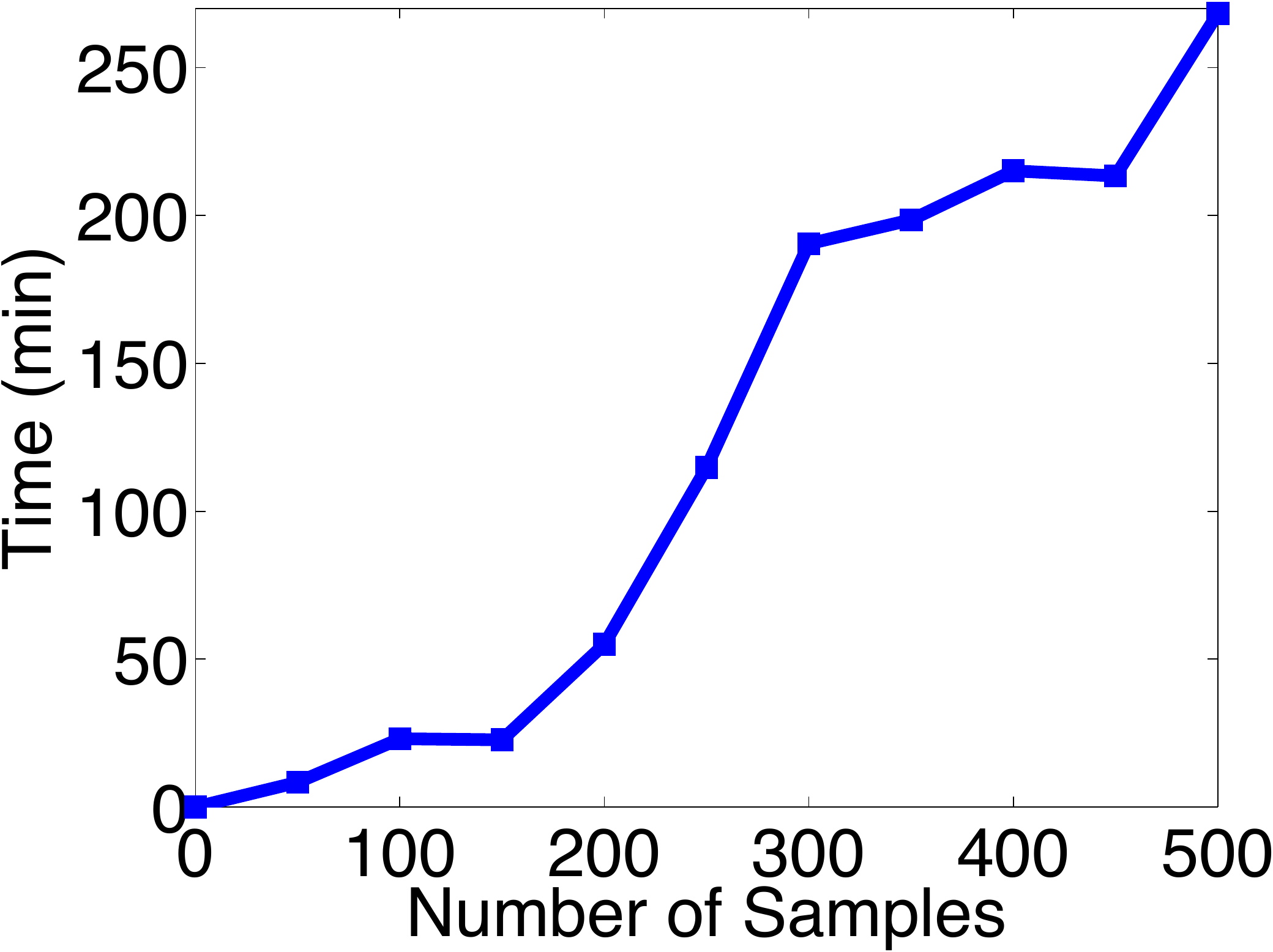}}
        \subfigure[Core-periphery]{\includegraphics[width=0.27\textwidth]{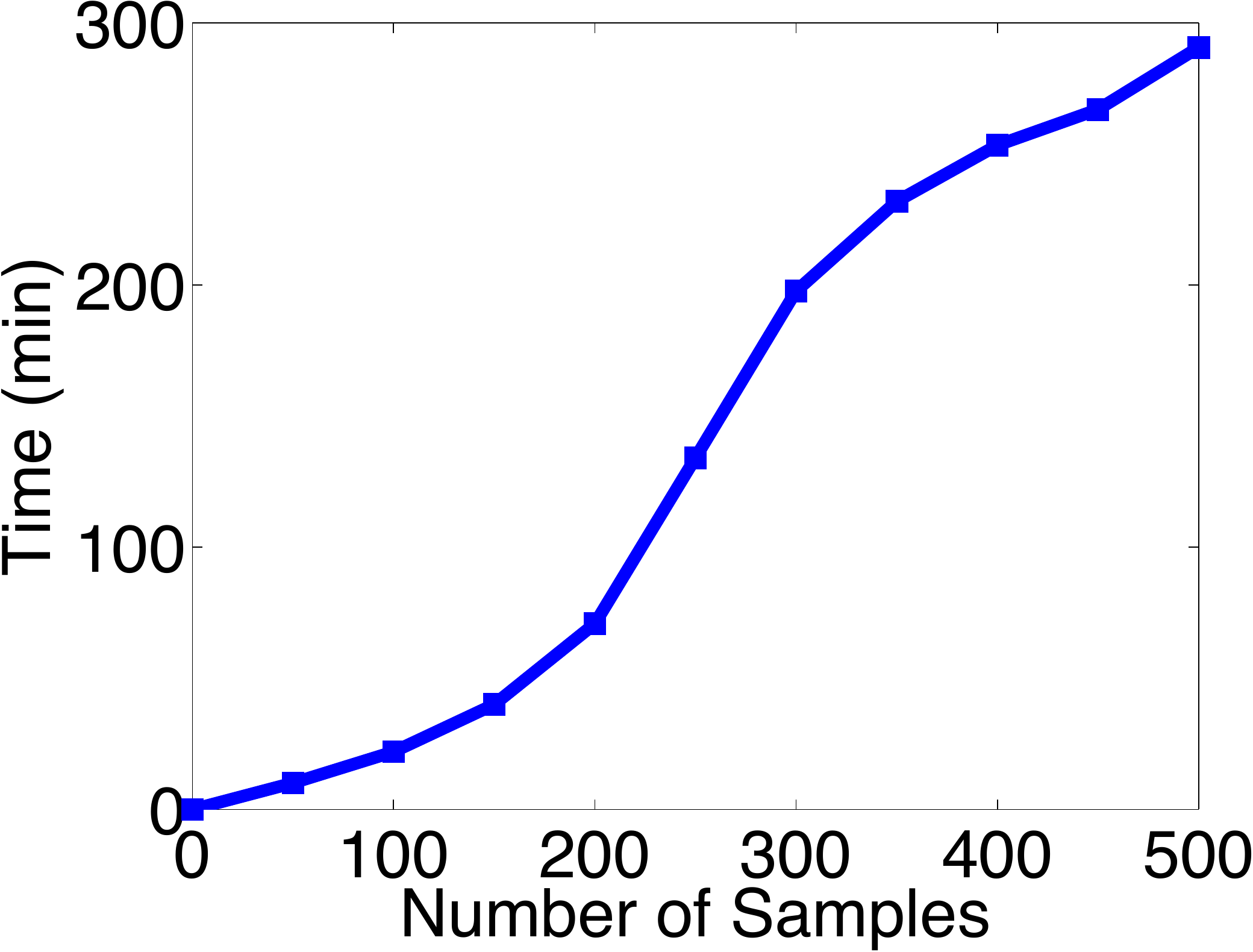}}
        \caption{Running time vs. number of samples.} \label{fig:time-vs-samples}
\end{figure*}

\xhdr{Running time vs. number of samples} 
Figure~\ref{fig:time-vs-samples} plots the average running time against the number of Montecarlo samples used to approximate the likelihood, Eq.~\ref{eq:sampling}.
%
%

\clearpage
\newpage

\xhdr{Toy example} We consider the same 64-node hierarchical Kronecker network as in Section~\ref{sec:experiments-synthetic} and vi\-sua\-lize the approximate likelihood given by 
Eq.~\ref{eq:sampling-simplified} against number of observed cascades ($C=1,\ldots,8$) for each node in the network using $150$ Monte Carlo samples.
\begin{figure*}[h]
        \centering
        \includegraphics[width=0.65\textwidth]{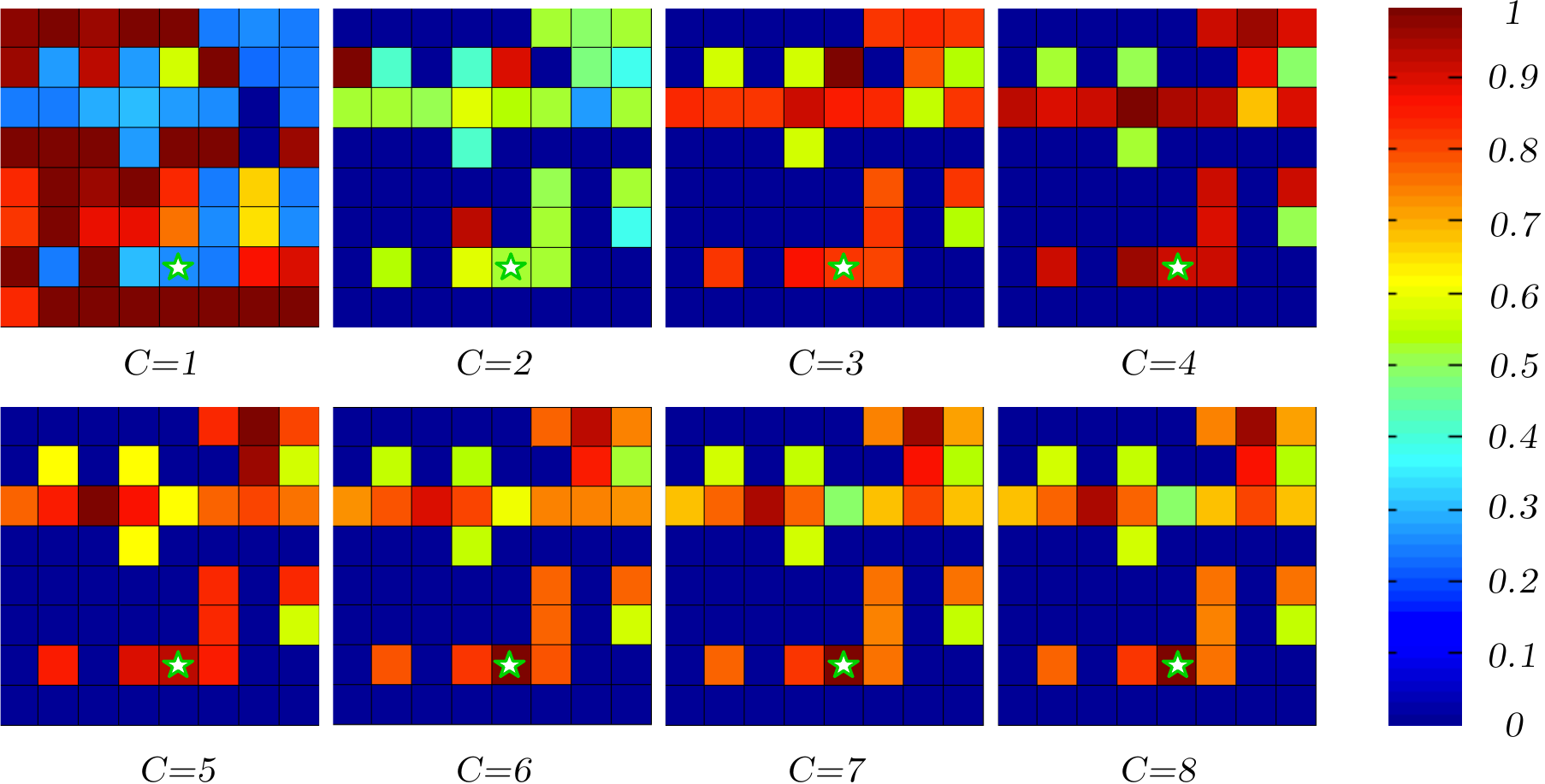} 
        \caption{Evolution of the proposed method with respect to the number of cascades.} \label{fig:grid-all}
\end{figure*}

\begin{figure*}[h]
        \centering
        \subfigure[SP]{\includegraphics[width=0.23\textwidth]{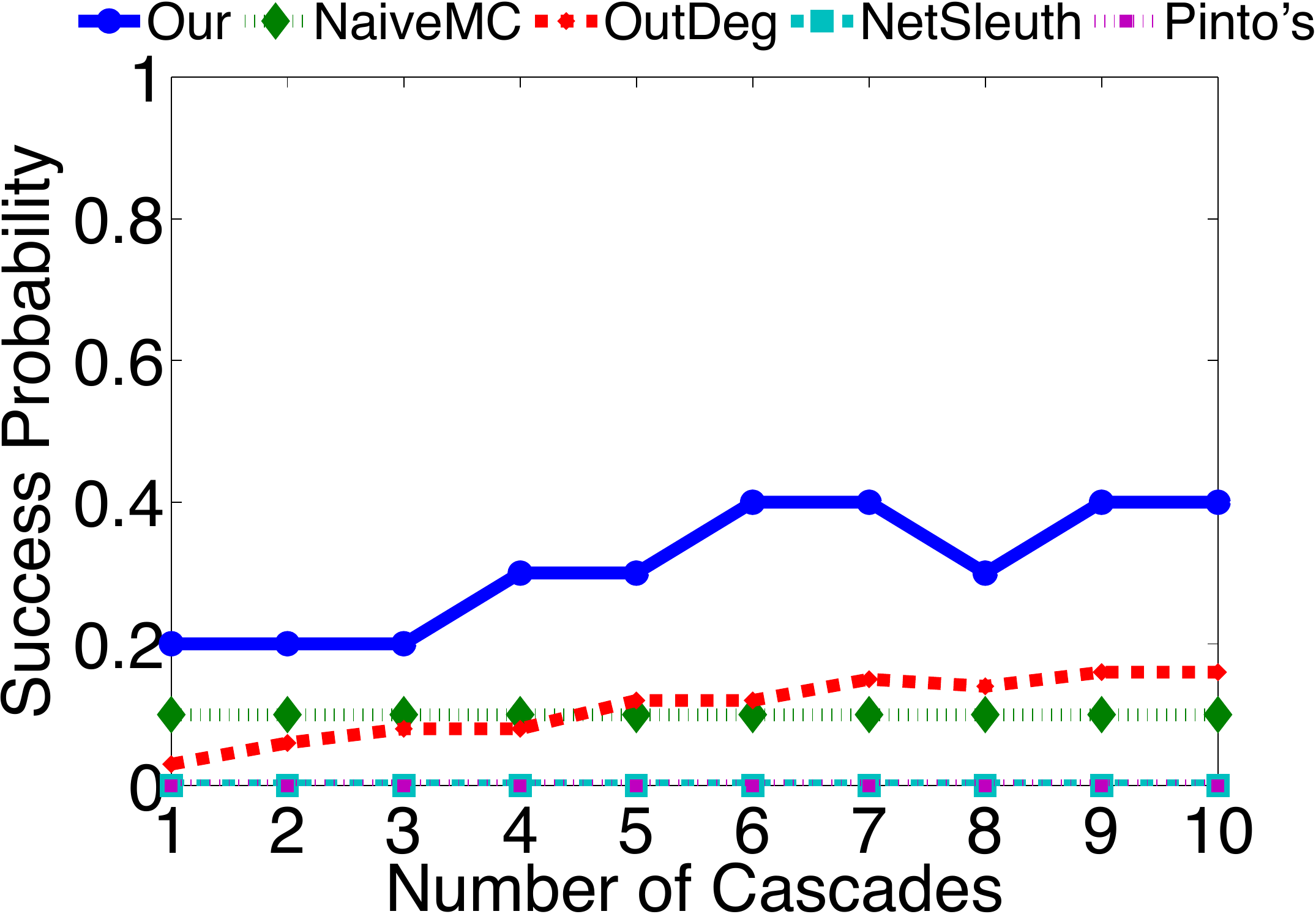}}
        \subfigure[Top-10 SP]{\includegraphics[width=0.23\textwidth]{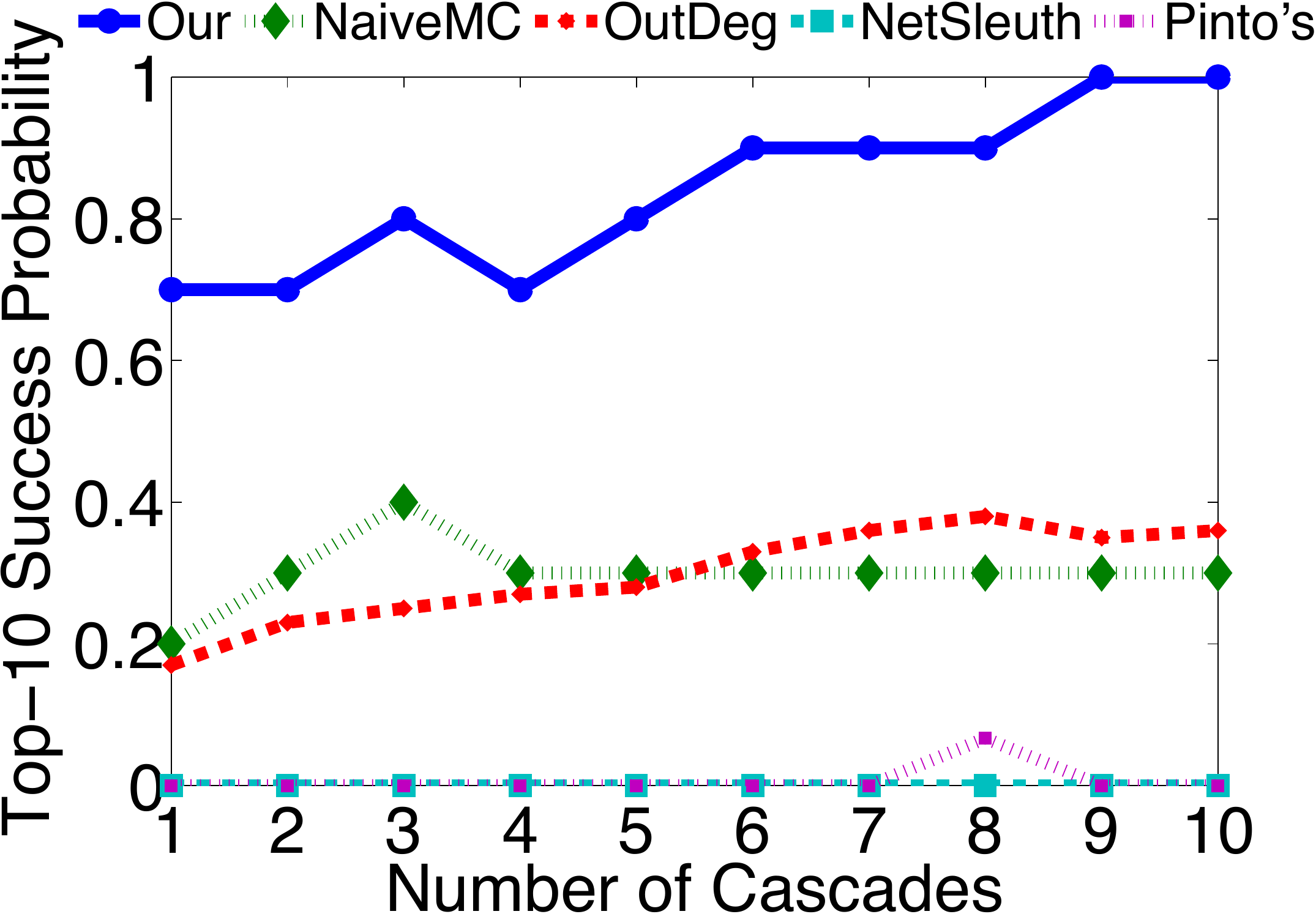}}
        \caption{Success Probability (SP), Top-10 Success Probability (Top-10 SP) and Mean-squared error (MSE) on the estimation of $t_s$ for a hierarchical Kronecker network.} \label{fig:acc-cascades-hierarchical}
\end{figure*}

\xhdr{Accuracy on a hierarchical Kronecker network} We additionally evaluate the accuracy of our method in comparison with the same two state of the art methods and two baselines
as in Section~\ref{sec:experiments-synthetic} in a Kronecker hierarchical network. Figure~\ref{fig:acc-cascades-hierarchical} shows the success probability (SP) and top-10 success 
probability, and mean squared error (MSE) on the estimation of $t_s$.
\end{appendix}

\end{document}